\documentclass[useAMS,usenatbib]{mn2e}
\usepackage{graphicx}
\usepackage{rotating}
\usepackage{aas_macros}
\usepackage{amsmath}
\usepackage{color}

\title[The Nature of UCDs] 
{The Ubiquity and Dual Nature of Ultra Compact Dwarfs}

\author[Norris $\&$ Kannappan]{Mark A. Norris$^{1}$\thanks{manorris@physics.unc.edu} \& 
Sheila J. Kannappan$^{1}$\thanks{sheila@physics.unc.edu}\\ 
$^1$ Dept. of Physics and Astronomy, University of North Carolina, Chapel Hill, CB 3255, Phillips Hall, Chapel Hill,\\
   NC 27599-3255, USA \\
}

\begin{document}

\date{Accepted 2010 ***. Received 2010 ***; in original form ***}

\pagerange{\pageref{firstpage}--\pageref{lastpage}} \pubyear{2007}

\maketitle

\label{firstpage}

\begin{abstract}
We present the discovery of several Ultra Compact Dwarfs (UCDs) located in field/group 
environments. Examination of these objects, plus literature UCDs, confirms the existence 
of two distinct formation channels for these compact stellar systems. We find that the UCDs 
we have discovered around the group elliptical NGC3923 (and most UCDs in general) 
have properties consistent with their being the most luminous members of the host galaxy's 
globular cluster (GC) system. As with GCs they are therefore likely to be the product of the 
most violent epochs of galaxy formation. We describe UCDs of this type as giant GCs 
(GGCs). In contrast, the UCD we have found associated with the isolated S0 NGC4546 is
clearly the result of the stripping of a nucleated companion galaxy. The young age 
($\sim$3.4~Gyr) of the UCD, the lack of a correspondingly young GC population, the apparently
short dynamical friction decay timescale ($\sim$0.5~Gyr) of the UCD, and the presence of 
a counterrotating gas disc in the host galaxy (co-rotating with respect to the UCD) together 
suggest that this UCD is the liberated nucleus remaining after the recent stripping of a 
companion by NGC4546. We infer that the presence of UCDs of either category (GGC's 
formed in major star forming events, or stripped nuclei formed in minor mergers) can provide
a useful probe of the assembly history of the host galaxy. We suggest a general scheme 
that unifies the formation of GCs, UCDs, and galaxy nuclei. In this picture ``normal" GCs are 
a composite population, composed of GCs formed \textit{in situ}, GCs acquired from accreted 
galaxies, and a population of lower mass stripped dwarf nuclei masquerading as GCs. Above 
a ``scaling onset mass" of 2$\times$10$^6$~M$_\odot$ (M$_{\rm V} \sim$ --10), UCDs 
emerge together with a mass-size relation and a likely mass-metallicity relation (the ``blue tilt"). 
In the mass range up to 7$\times$10$^7$~M$_\odot$ (M$_{\rm V} \sim$ --13) UCDs comprise 
a composite population of GGCs and stripped nuclei. Interestingly, dwarf nuclei have similar 
colours to blue GCs and UCDs across the scaling onset mass, smoothly extending the blue tilt, 
while nuclei of more massive galaxies and a prominent minority of UCDs extend the red locus 
of GCs. Above 7$\times$10$^7$~M$_\odot$, UCDs must be almost exclusively stripped nuclei,
as no sufficiently rich GC systems exist to populate such an extreme of the GCLF.

\end{abstract}


\section{Introduction}
\label{sec:intro}

Over the past decade increasing numbers of an enigmatic class of stellar system with 
luminosities and sizes intermediate between globular clusters (GCs) and dwarf or 
compact elliptical galaxies (dEs or cEs) have been discovered. Since their discovery 
in the Fornax cluster \citep{Hilker99,Drinkwater00} these ultra-compact dwarfs 
\cite[UCDs,][]{Phillips01} have been found in the cores of other galaxy clusters such 
as Virgo \citep{Hasegan05,Jones06}, Hydra I \citep{Wehner07,Misgeld08}, Coma 
\citep{Price09,Madrid10}, and Centaurus \citep{Mieske09}. More recently the first UCDs 
found outside of clusters have been discovered. The first confirmation was of one 
definite and four possible UCDs in a survey of six galaxy groups by \cite{Evstigneeva07b}, 
and the second was the spectroscopic confirmation by \cite{Hau09} of a single UCD 
associated with the Sombrero galaxy (NGC4594/M104), which resides in a field 
environment. Subsequently \citet{DaRocha10} discovered a total of 13 UCDs in 
two Hickson Compact Groups.

At present the information provided by the properties of UCDs is often somewhat
contradictory, leading to the consideration of three main hypotheses for the formation 
of UCDs:
 
(1) That UCDs are simply very massive ``normal" GCs \citep[e.g.,][]{Drinkwater00,Mieske02},
formed from the collapse of individual giant molecular clouds, possibly influenced by
distinct processes such as self shielding that emerge at higher mass \citep{Murray09}.

(2) That UCDs (plus some of what are currently considered massive GCs) are formed 
by the merger of young massive star clusters (YMCs) produced during the most violent 
epochs of galaxy formation such as during galaxy-galaxy mergers 
\citep{Fellhauer02,Maraston04}. YMCs that did not undergo merging would be left
to evolve into the ``normal" GC population of the resulting galaxy.

(3) That UCDs are the remnant nuclei of nucleated galaxies that
have been stripped of their envelopes by tidal interaction \citep{Bekki01,Bekki03b}.

In both scenarios (1) and (2) the UCD hosting galaxy is expected to harbour a 
population of GCs with ages and abundances nearly identical to those of the UCDs 
with which they formed. Similarly, in both cases the luminosity function of UCDs 
associated with a particular galaxy should be strongly correlated with the GC luminosity 
function of that galaxy, irrespective of host galaxy environment. Discriminating between 
objects formed through route (1) or (2) will depend on the ability to predict and observe 
differences in the structural properties of UCDs formed by each route; we refer to both 
as giant globular clusters (GGCs). As both scenarios are intimately tied up with globular 
cluster formation, which is thought to occur mainly during the most violent periods of 
galaxy assembly, GGC-type UCDs can serve as tracers of the major epochs of galaxy 
formation.

In contrast, the third UCD formation scenario is related to minor galaxy mergers. 
Unlike in (1) and (2), in a stripping scenario there is no expectation of correlations 
between the stellar populations and luminosity functions of the UCDs and those of 
any GCs present. The stellar populations of UCDs formed by stripping may also  
display environmental dependences. Stripping should, on average, occur later in 
lower density environments, due to the hierarchical nature of structure formation. 
The nuclei of galaxies in lower density environments are also observed to display 
a range of ages, in contrast to the uniformly old nuclei found in the densest regions 
of galaxy clusters \citep{Paudel10b,Paudel10a}. This should lead to younger average 
ages for UCDs in lower density environments. Additionally the total number of UCDs 
formed by stripping is expected to be larger, and the UCDs spread more widely, in 
higher density environments where the tidal field is stronger \citep{Bekki03c}. These 
stripped-nucleus UCDs could also display significantly higher mass-to-light ratios, 
relative to GGC UCDs, due to the lingering presence of a dark matter component not 
found in GCs (\citealt{Goerdt08}, although see \citealt{Bekki03c} for an alternative view).

The lack of consensus that a single scenario provides a satisfactory explanation for the 
properties of all known UCDs has led to the suggestion that UCDs as a class consist
of a ``mixed bag" of objects  \citep{Hilker09b,Taylor10,DaRocha10} formed by different 
routes. Only by increasing the sample size and range of parameter space probed by 
UCD studies can we hope to determine which formation routes are at work. One area 
of parameter space currently undersampled is the behaviour of UCD properties such 
as stellar populations, structure, and frequency as environment changes from dense 
cluster core to the field.

In the this paper, we analyse imaging and spectroscopy of several newly discovered 
UCDs associated with the group shell elliptical NGC3923 and the field S0 galaxy NGC4546, 
plus the known UCD near the Sombrero, a field Sa galaxy, as part of a small pilot survey 
of UCDs in field/group environments. We then compare to literature data to assess UCD 
formation scenarios. The paper is structured as follows: In Section 2 we describe the data 
obtained, its reduction, and analysis. In Section 3 we present our results. In Section 4 we 
provide a discussion of the implications of our results. In Section 5 we provide some 
concluding remarks.

\begin{table*}
\begin{center}
\begin{tabular}{lcccc} \hline
Parameter 				& 	NGC3923 		& 	NGC4546			&	Sombrero			&	 Units		 	\\
\hline
R.A.						&	11:51:01.8		& 	12:35:29.5		&	12:39:59.4		& 	h:m:s (J2000) 		\\
Dec.						&	$-$28:48:22		& 	$-$03:47:35           	&	 $-$11:37:23			&       d:m:s (J2000)   		\\
V$_{\rm hel}$				&     	1739$\pm$9    		& 	1050$\pm$9		&	1024$\pm$5		&  	kms$^{-1}$		\\
Type						&	E4-5				&	SB(s)0-			&	SA(s)a			& 	 -				\\
A$_{\rm B}$				&	0.357			&	0.146			&	0.221			&	mag				\\	
m-M$^{a}$				&  	31.64$\pm$0.28	& 	30.58$\pm$0.20     	&	29.77$\pm$0.03 	& 	mag                  		\\
Distance$^{b}$				&   	21.28$\pm$2.93	&	13.06$\pm$1.26	&	9.0$\pm$0.1		& 	Mpc                  		\\
Scale$^{b}$				&	103				&	63				&	44				&	pc/arcsec			\\
Stellar Mass$^{c}$			&	1.7$\times$10$^{11}$				&	2.7$\times$10$^{10}$				&	8.2$\times$10$^{10}$				&	M$_\odot$		\\
\hline
\end{tabular}
\end{center}
\caption[Basic Properties]{Properties of NGC3923, NGC4546 and the Sombrero. From NED (http://nedwww.ipac.caltech.edu) unless otherwise noted.\\
$^{a}$ From \citet{TonrySBF4} for NGC3923 and NGC4546 after applying $-$0.16 correction for
recalibration of the surface brightness fluctuations distance scale by \cite{Jensen03}. The Sombrero distance is from 
\citet{Spitler06}.\\ 
$^{b}$ Calculated from m-M. $^{c}$ Calculated as described in Section \ref{sec:stellar_mass_estimates}.}
\label{tab:galaxy_properties}
\end{table*}

\section{Observations and Data Reduction}

In this section we describe the data used in this study, as well as its subsequent 
analysis. We use archival \textit{HST} imaging to select UCD candidates (plus 
comparison GCs for NGC3923) and follow-up SOAR spectroscopy to confirm the 
redshift association of one UCD with NGC4546, plus two UCDs with NGC3923. 
A third UCD candidate near NGC3923 has not been spectroscopically observed 
to date due to its lower luminosity. The description of the reduction of the data is 
straightforward, so those less interested in the finer points of the data analysis may 
wish to read Section \ref{sec:target_selection} only (on the sample) before continuing 
to Section \ref{Sec:results}.

\subsection{UCD Identification}
\label{sec:target_selection}

Intrigued by the possibility that UCDs are common in all environments,  we searched 
archival \textit{HST} ACS or WFPC2 observations of non-cluster galaxies for UCDs. 
This search consisted of visually examining all \textit{HST} ACS or WFPC2 
observations of bright galaxies (M$_{\rm B}$$<$~$-$19.0) at a distance of 10 to 30~Mpc 
observable with the SOAR telescope during spring (6$<$R.A.$<$18hrs, 
$-$70$<$Dec$<$0 deg), 76 galaxies in all. In total 11 UCD candidates were found 
(including the previously confirmed Sombrero UCD), associated with nine galaxies. 
Here we focus on a subset of the UCD candidates for which we have obtained 
spectroscopic confirmation; the remaining objects will be included in an upcoming 
paper examining the entire HST archive for UCDs.

UCD candidates associated with the group shell elliptical NGC3923 and the isolated 
S0 NGC4546 were selected for spectroscopic follow-up. These targets were selected 
because of the availability of ancillary data for the target galaxies: high S/N Gemini/GMOS 
MOS spectroscopy of  NGC3923 GCs from \citet{Norris08}, and SAURON IFU spectroscopy 
of NGC4546 from \citet{Kuntschner06}. We also include the Sombrero UCD in the following 
analysis, making use of the properties derived by \citet{Hau09}.

Table \ref{tab:galaxy_properties} displays the basic properties of the galaxies under study, 
including the assumed extinction, distances and projected scale. Table \ref{tab:observing_log} 
provides the log of observations, including our SOAR/Goodman imaging and spectroscopy 
and \textit{HST} observations. Table \ref{tab:ucd_photometry} presents the derived photometry 
and recessional velocities of our UCDs. The unconfirmed UCD in the NGC3923 system is 
labeled ``UCD3c" to indicate that it is still a candidate.

\begin{table}
\begin{center}
\begin{tabular}{lccc} \hline
Galaxy						& Date		& 	Exposure Time						&	 Seeing	\\
\hline
NGC3923 spec$^{\star}$			& 15/04/09	& 	8$\times$1200s					& 	0.6$''	$	\\
NGC4546 spec$^{\star}$			& 18/04/09	& 	6$\times$1200s					& 	0.5$''$		\\
NGC3923 im$^{\star}$			& 13/05/09	& 	4$\times$60s in B,V$\&$R			& 	0.8$''$		\\
NGC4546 im$^{\star}$			& 13/05/09	& 	4$\times$60s in B,V$\&$R			& 	1.3$''$		\\
Sombrero im$^{\star}$			& 13/05/09	& 	4$\times$60s in B,V$\&$R			& 	1.4$''$		\\
NGC3923 ACS$^{\dagger}$		& 07/12/02	&	1140$^{a}$, 978s$^{b}$				&	-	\\
NGC4546 WFPC2$^{\diamond}$	& 16/05/94	&	160s$^{a}$						&	-	\\
Sombrero ACS$^{\ddagger}$		& 08/06/03	&	2.7$^{c}$, 2.0$^{d}$, 1.4ks$^{e}$		&	-	\\
\hline
\end{tabular}
\end{center}
\caption[Observing Log]{Log of Observations\\
$^{\star}$ SOAR/GHTS $^{\dagger}$ Prop ID:9399 PI:Carter  \\
$^{\diamond}$ Prop ID:5446 PI:Illingworth  $^{\ddagger}$ Prop ID:9714 PI:Noll. \\ 
$^{a}$ F606W$\sim$V.
$^{b}$F814W$\sim$I. 
$^{c}$F435W$\sim$B.
$^{d}$F555W$\sim$V.
$^{e}$F625W$\sim$R.}
\label{tab:observing_log}
\end{table}

\subsection{SOAR Spectroscopy}
\label{sec:soar_spectroscopy}

Spectroscopic exposures of two NGC3923 UCD candidates were taken on 2009 April 15 
during commissioning of the MOS mode of the Goodman High Throughput Spectrograph 
\citep[GHTS,][]{GoodmanSpec} on the 4.1m Southern Astrophysics Research (SOAR) 
Telescope. These observations comprised 8$\times$1200s exposures through 1.5$''$ wide 
MOS slitlets, dispersed with a 600l/mm grating, providing a spectral resolution of $\sim$6.2~\AA. 
The proximity of the UCD candidates to NGC3923 (R$<$8~kpc) meant that the galaxy diffuse 
light contributed significant flux to each slitlet. Therefore, to allow the accurate ``sky" subtraction 
of each UCD spectrum the slitlets were long ($>$17$^{\prime\prime}$). Both UCDs have 
spectroscopic coverage between 4520 and 7075~\AA, including coverage of the H$\beta$, 
Mg$b$, Fe5270, and Fe5335 absorption features.

Spectroscopic observations of the NGC4546 candidate UCD were undertaken on 2009 April 18 
with the Goodman spectrograph in longslit mode. The 1.68$''$ wide longslit was aligned such 
that the target UCD and the centre of NGC~4546 both fell on the longslit (PA: 140.7$^\circ$). 
The object was exposed for 6$\times$1200s, again utilising the 600~l/mm grating, this time 
providing a spectral resolution of $\sim$6.3~\AA\, and a wavelength coverage of 4350 $-$ 7000~\AA.

Standard $\textsc{IRAF}$ routines were used to carry out bias subtraction, flatfielding, and 
wavelength calibration. Spectra from individual exposures were traced and extracted using 
$\textsc{apall}$. Because of the relatively large regions available for sky estimation and the 
varying contribution of host galaxy diffuse light across the slit/slitlets, the background ``sky" 
across each spectral column was fitted with a low order polynomial and subtracted. The 
extractions were unweighted as the use of variance weighting was inadvisable due to the 
slightly resolved nature of these objects. Individual extracted spectra were combined using 
the task $\textsc{scombine}$.

\begin{figure} 
   \centering
   \begin{turn}{0}
   \includegraphics[scale=1.00]{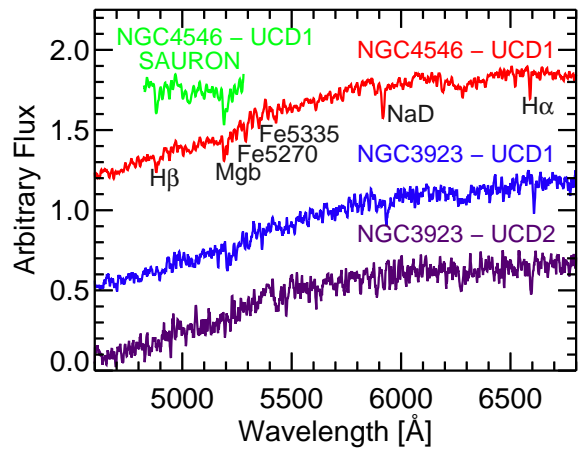}
   \end{turn} 
   \caption{SOAR/Goodman UCD spectra plus SAURON continuum-normalised spectrum 
   of NGC4546 UCD1 (full coverage shown). Redshifted absorption lines are apparent.}
   \label{fig:UCD_Spectra}
\end{figure}

The final UCD spectra are displayed in Figure \ref{fig:UCD_Spectra}. Redshifted absorption 
lines are obvious in NGC4546-UCD1 and NGC3923-UCD1, and the resulting spectra are of 
sufficient S/N to measure absorption line velocities in all cases. However, due to a since rectified 
problem with stray light sources within the instrument, the spectra are not suitable for the study 
of the stellar populations of  the UCDs. It proved impossible to adequately remove the stray light 
signal during processing, leading to compromised line strength estimates.

Recessional velocities were measured using the IRAF Fourier cross correlation code \textsc{fxcor}. 
We made use of MILES simple stellar population models \citep{Vazdekis10} as velocity templates, 
using 24 SSPs spanning the age range 1.8 to 12.6~Gyr and the metallicity range [M/H] = $-$1.68 
to +0.2, to minimise the effects of template mismatch. The quoted velocity in Table \ref{tab:ucd_photometry} 
is the 3$\sigma$ clipped mean of the 24 determined velocities (one per template), quoting the 
median error for the unclipped velocities.

\subsection{SAURON Spectroscopy}
\label{sec:Sauron_spec}

\begin{figure} 
   \centering
   \begin{turn}{0}
   \includegraphics[scale=0.52]{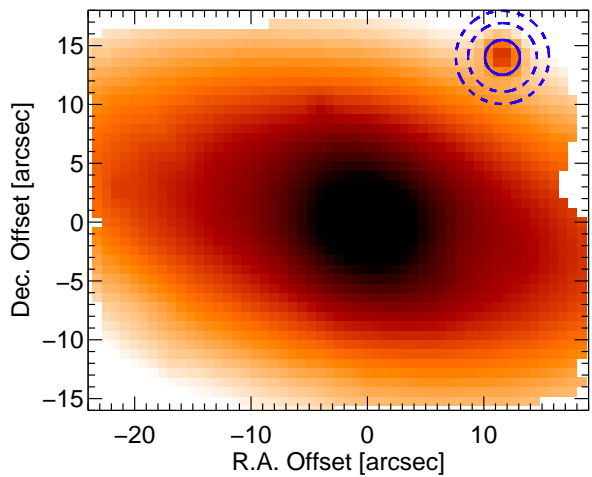}
   \end{turn} 
   \caption{Log(Flux) in the SAURON NGC4546 datacube. The UCD is obvious in the
   top right. The solid blue circle shows the radius within which the lenslets are
   coadded to produce the UCD spectrum. The dashed blue circles denote the
   annulus within which lenslets are coadded to produce an estimate of the sky + 
   NGC4546 diffuse light background.}
   \label{fig:SAURON_datacube}
\end{figure}

To investigate the stellar populations of the NGC4546 UCD, and to allow a direct comparison 
to NGC4546 itself, we make use of SAURON IFU observations \citep{SAURON1}. The SAURON 
datacube provides an on-sky coverage of 42.4$^{\prime\prime}$$\times$33.6$^{\prime\prime}$ 
sampled with 0.94$^{\prime\prime}$ lenslets, Figure \ref{fig:SAURON_datacube} displays the 
log(flux) of the datacube (determined by summing the flux in each spatial element), with the UCD 
visible towards the top right. To produce the UCD spectrum, spatial elements within 1.5$^{\prime\prime}$ 
of the centre of the UCD were coadded (12 in all). The sky, including NGC4546 diffuse light, was 
determined by averaging spatial elements within an annulus from 3 to 4$^{\prime\prime}$ (36 in all). 
The final UCD spectrum covers the wavelength range 4825 to 5280~\AA\, with a FWHM of $\sim$4.4~\AA.

The recessional velocity of the NGC4546 UCD was determined using the same procedure as in 
Section \ref{sec:soar_spectroscopy}. The measured velocity is consistent with that determined from
our lower S/N SOAR spectroscopy. We also used the IDL code pPXF \citep{ppxf} to provide an 
additional check of the UCD velocity and to confirm that the $\sigma$ of the UCD is too low to 
measure (i.e., $<$ 50~kms$^{-1}$).

To derive stellar population information from the UCD spectrum we measure Lick/IDS absorption 
line indices following the procedure outlined in \citet{Kuntschner06}. We degrade the SAURON 
spectrum with FWHM$\sim$4.4~\AA\, to the resolution of the Lick/IDS system in the same wavelength 
range ($\sim$8.4~\AA). We do not correct for the line of sight velocity distribution of the UCD, as its 
$\sigma$ is less than 100~kms$^{-1}$ and the correction is therefore negligible \citep[][]{Kuntschner04}. 
Due to the limited wavelength coverage we can measure only three line indices, H$\beta$, Fe5015 
and Mg$b$, using the standard definitions of \citet{Trager98}. The last step required to place our 
measured indices on the Lick/IDS system is to correct small systematic offsets due to continuum 
shape differences between the SAURON and Lick/IDS spectra. To do this we use the offsets of 
\citet[][see \citealt{NSK06} for more details]{Kuntschner06}.

\subsection{SOAR Imaging}
\label{Sec:Soar_Imaging}

\begin{figure*} 
   \centering
   \begin{turn}{0}
   \includegraphics[scale=1.00]{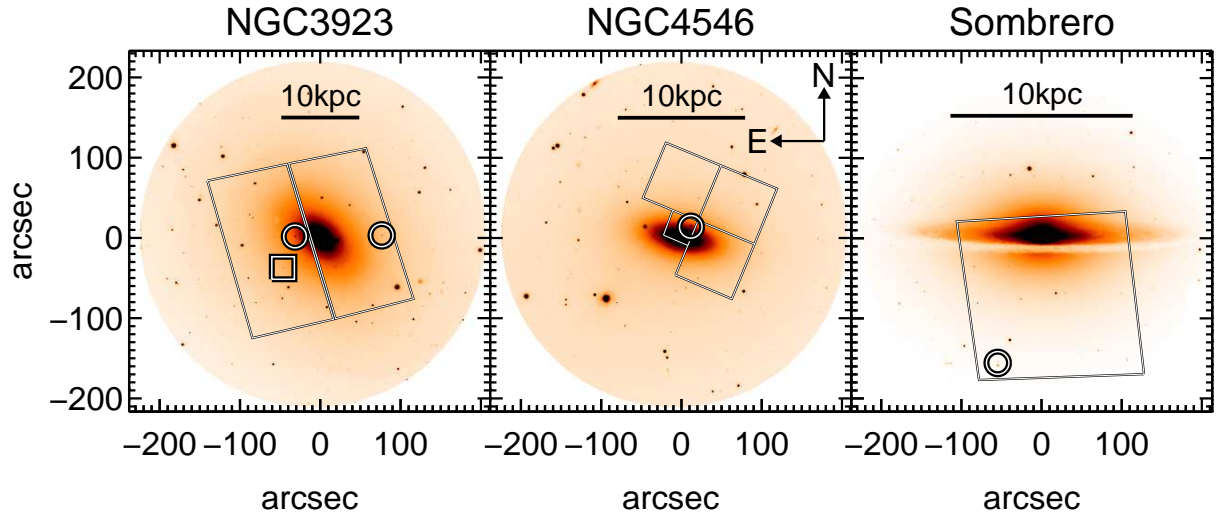}
    \end{turn} 
    \caption{SOAR/Goodman V-band images showing the location of our confirmed field/group 
    UCDs (circles) as well as the candidate third NGC3923 UCD (square), relative to their host 
    galaxies. The \textit{HST} ACS or WFPC2 footprints are also indicated. The solid black line 
    in each panel denotes 10~kpc at the assumed distance of the galaxies.}
   \label{fig:fieldofview}
\end{figure*}

We carried out BVR band imaging of NGC3923 and NGC4546 using the Goodman spectrograph 
in imaging mode. As a consistency check of our photometry, we also obtained SOAR/Goodman 
BVR imaging of the Sombrero UCD, which has previously published BVR imaging converted from 
the \textit{HST} equivalent bands by \cite{Hau09}. Coadded images with 4$\times$60s exposure 
time are shown in Figure~\ref{fig:fieldofview}. The imaging was binned 2$\times$2 providing a pixel 
scale of 0.291$''$, with seeing of $\sim$0.8 to 1.4$''$. 

In all cases the individual exposures were bias subtracted, flatfielded, and combined in $\textsc{IRAF}$ 
using standard routines. As all of our UCD candidates are located within 3$'$ of their host galaxies, 
care must be taken to remove the galaxy halo before accurate photometry can be achieved for the 
UCDs. To model the galaxy halos (and the sky background) the \textsc{IRAF} task $\textsc{median}$ 
was used to produce a median image, where each pixel was the median value of a 75$\times$75 
pixel box. The box size was chosen to be sufficiently large (21.8$''$$\times$21.8$''$) that the UCDs 
themselves did not affect the determination of the median. More details of this procedure and the 
method used to determine the optimum median box smoothing size are provided in Section \ref{Sec:HST_IP}.

After subtraction of the median image, the photometric measurements were carried out with 
\textsc{SExtractor} \citep{Sextractor96}. The same procedure was used to reduce observations of the 
standard star field \citep[SA104,][]{Landolt09} observed prior to the galaxy imaging. Five standard stars 
were used to calculate the photometric zeropoints, providing zeropoint uncertainties of $<$~0.03~mag 
in all bands. Finally all photometry was corrected for Galactic extinction using the extinction maps of \cite{Schlegel1998}.

To ensure that our photometry was accurately placed on standard photometric systems we compared 
our \textsc{SExtractor} output catalog of Sombrero galaxy GCs/UCD with the \textit{HST} ACS photometry 
previously presented for these objects in \cite{Spitler06} and \cite{Hau09}. In all three bands systematic
offsets are $<$0.03~mag, validating the accuracy of our derived zeropoints.

\subsection{IR Imaging}
The UCDs of NGC4546 and the Sombrero are sufficiently luminous that they are detected in J, H, 
and K$_{\rm s}$ by 2MASS  \citep{Skrutskie06}. For the UCDs associated with NGC3923, we 
have made use of archival K$_{\rm s}$ imaging from the SOFI instrument on the 3.6 m ESO New 
Technology Telescope. These data have previously been described in \cite{Brown03}. Briefly, 
they comprise 30$\times$30s exposures with a pixel scale of 0.292 arcsec pixel$^{-1}$ and a 
field of view of 4.94$\times$4.94 arcmin$^{2}$. All three NGC3923 UCD candidates are located 
within this field of view. We reduced the data in $\textsc{IRAF}$ using standard routines to subtract 
the dark, produce and apply a flat field constructed from the object frames, combine the individual 
exposures with a sigma clip, and carry out a median image subtraction (box size 50$\times$50 pixels) 
to remove the galaxy halo and sky. The zero point of the image was determined by comparison 
with 2MASS photometry of 9 isolated stars within the FOV. Finally, a small correction for Galactic 
extinction was applied based on the dust extinction maps of \cite{Schlegel1998}.

\subsection{HST Imaging}
Table \ref{tab:observing_log} lists the \textit{HST} imaging available for our three target galaxies.
For NGC4546 and the Sombrero galaxy we make use of \textit{HST} imaging only to derive structural 
properties of the UCDs. This is because only one band of \textit{HST} imaging is available for 
NGC4546 and because the photometric properties of the Sombrero UCD have been extensively 
discussed previously in \cite{Hau09}.

The GC system of NGC3923 has previously been examined with \textit{HST} ACS Wide Field 
Camera photometry comprising 1140s in F606W ($\sim$V) and 978s in F814W ($\sim$I). The 
ACS WFC has a pixel scale of 0.05$^{\prime\prime}$, which at the distance of NGC3923 equates 
to 5.15 pc/pixel. These data have been described in detail in \cite{Sikkema06} and \cite{ChoThesis}. 
However, in both cases the UCD candidates appear to have been ignored, presumably because 
they are clearly resolved and their luminosities are too high to be standard GCs. We therefore 
re-examined these data to provide accurate V and I band photometry as well as structural information 
for our candidate UCDs and the GCs of NGC3923.

\subsubsection{Background Subtraction}
\label{Sec:HST_IP}

\begin{figure} 
   \centering
   \begin{turn}{0}
   \includegraphics[scale=1.00]{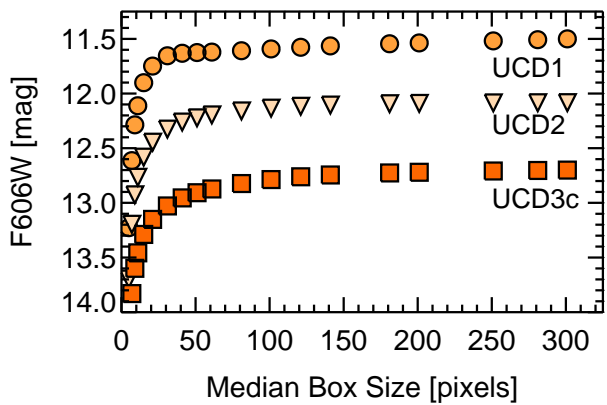}
   \end{turn} 
   \caption{Demonstration of how the choice of median smoothing box size affects the total magnitude 
   summed within 250 pc (48.5 pixels) of our NGC3923 UCDs. A box size of 251$\times$251 pixels is 
   sufficiently large to be unaffected by the presence of the UCDs.}
   \label{fig:Smooth_Test}
\end{figure}

\begin{figure*} 
   \centering
   \begin{turn}{0}
   \includegraphics[scale=0.95]{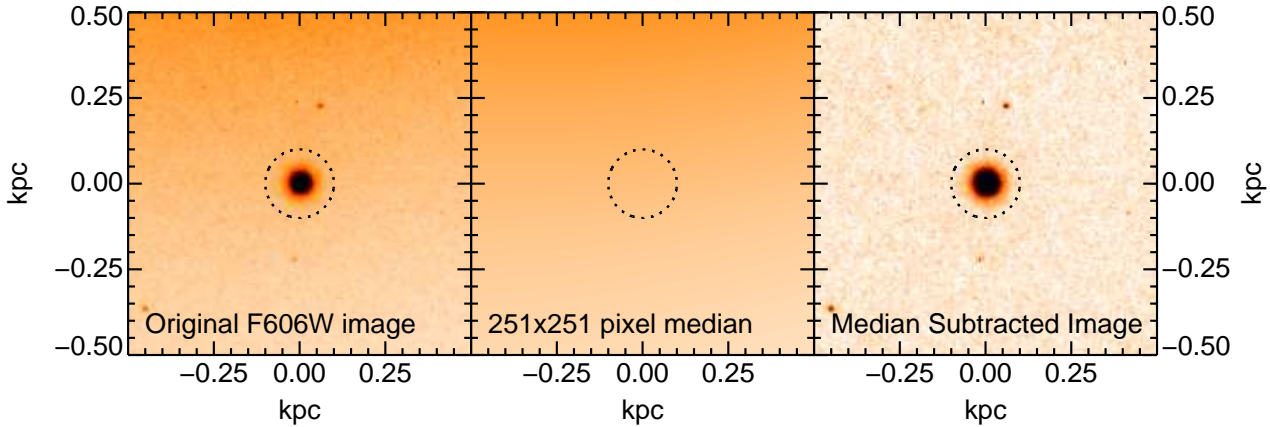}
   \end{turn} 
   \caption{Demonstration of the median subtraction of the NGC3923 halo. The left panel is the input 
   ACS F606W image, centred on NGC3923 UCD1. The central panel is the 251$\times$251 pixel 
   median image. The right panel is the median subtracted image used to measure UCD photometry.
   The dashed circles in all three panels show the maximum extent of UCD1 determined by a curve of 
   growth analysis.}
   \label{fig:Smooth_Test2}
\end{figure*}

The determination of integrated magnitudes and especially structural parameters for GCs and 
UCDs depends sensitively on the careful removal of the host galaxy halo light.  We have therefore 
applied the same median background removal approach mentioned in Section \ref{Sec:Soar_Imaging}
to the \textit{HST} data as well. The choice of median box size for the smoothing is particularly 
critical. Too small a box size results in the subtraction of UCD flux, while too large a box size results 
in an inaccurate removal of the host galaxy halo. We therefore investigated the optimum choice of 
median box size for our imaging. To do this for NGC3923, for example, we produced median images
of the F606W imaging of NGC3923 with a range of box sizes of between 5$\times$5 and 301$\times$301 
pixels (25.75$\times$25.75 to 1550$\times$1550~pc). We then subtracted these smoothed models
from the original image and measured the total magnitude of each of our three NGC3923 UCD 
candidates contained within a 250~pc radius (curve-of growth analysis shows that no UCD luminosity 
is expected beyond this radius, see next section). Figure \ref{fig:Smooth_Test} displays the result of 
this analysis. The total magnitude of each UCD converges to a constant value for box smoothing sizes 
of 251 pixels or more. To determine the size of sky residuals left by each smoothing box we also
measure the residual sky counts in a region adjacent to each of the three UCD candidates. For all 
three NGC3923 UCD candidates a 251$\times$251 pixel smoothing box offers the optimum choice
of box size, providing the correct total UCD flux and lowest sky residuals. The corresponding values 
for the NGC4546 and Sombrero UCDs are similar at 201$\times$201 and 301 $\times$ 301 pixels 
respectively. We caution that this fortunate result will not hold true in all cases: the optimum choice 
of smoothing box depends on several factors, including UCD size and structure as well as the behaviour 
of the galaxy background and the proximity of the object to the edges of the chip. In general for UCDs 
located near to galaxies it is necessary to determine the optimum smoothing scale for each UCD 
candidate individually. Figure \ref{fig:Smooth_Test2} displays an example of the result of this approach 
for the region around NGC3923-UCD1; as can be seen the galaxy halo is very effectively removed.

\subsubsection{Integrated Magnitudes}
\label{Sec:HST_IM}

We used the curve of growth method to determine total magnitudes, applying extinction corrections 
following \citet{Schlegel1998} and converting between HST and Johnson-Cousins filter systems 
using the method of \cite{Sirianni05} with a G2V spectral energy distribution. The \textit{HST} derived 
magnitudes are given in Table~\ref{tab:ucd_photometry} and agree with SOAR photometry where 
available. As only single-band \textit{HST} photometry is available for NGC4546 we do not convert 
its \textit{HST} photometry to the standard Johnson-Cousins system. Our ground based photometry 
is therefore used in all analyses of the NGC4546 UCD that rely on integrated magnitudes. For 
analyses of integrated magnitudes involving the Sombrero UCD we make use of the photometry 
obtained by \cite{Hau09}. We make no correction for internal extinction by the host galaxies, as none 
of our UCD candidates are located near obvious dust features.

\begin{table*}
\begin{center}
\begin{tabular}{lcccccccc} \hline
Name			&	R.A.			&	Dec.			&	B				&	V					&	R				&	I					&	K$_{\rm s}$			&	V$_{helio}$ 		\\
				&	[h:m:s]		&	[d:m:s]		&	[mag]			&	[mag	]				&	[mag]			&	[mag]				&	[mag]			&	[kms$^{-1}$]		\\
\hline
NGC3923 UCD1	&	11:51:04.1	&	--28:48:19.8	&	20.08$\pm$0.10	&	19.21$\pm$0.04$^*$	&	18.73$\pm$0.04	&	18.10$\pm$0.04$^*$	&	16.26$\pm$0.06	&	2096.9$\pm$17.8	\\
NGC3923 UCD2	&	11:50:55.9	&	--28:48:18.4	&	20.49$\pm$0.12	&	19.71$\pm$0.05$^*$	&	19.32$\pm$0.05	&	18.70$\pm$0.05$^*$	&	17.10$\pm$0.10	&	1500.5$\pm$44.2	\\
NGC3923 UCD3c	&	11:51:05.2	&	--28:48:58.9	&	21.18$\pm$0.16	&	20.35$\pm$0.07$^*$	&	19.95$\pm$0.07	&	19.38$\pm$0.07$^*$	&	18.07$\pm$0.17	&	-				\\
NGC4546 UCD1	&	12:35:28.7	&	--03:47:21.1	&	18.57$\pm$0.05	&	17.64$\pm$0.04		&	17.15$\pm$0.03	& 		-				&	14.86$\pm$0.17	&	1255.9$\pm$24.0	\\
Sombrero UCD1	&	12:40:03.1	&	--11:40:04.3	&	18.37$^*$			&	17.46$^*$				&	16.88$^*$			& 		-				&	14.70$\pm$0.11			&	1293.1$\pm$9.5	\\
\hline
\end{tabular}
\end{center}
\caption[UCDs]{Basic properties of four confirmed UCDs and one more UCD candidate in the three 
systems studied. B,V,R and I magnitudes are from SOAR/Goodman or \textit{HST} imaging (\textit{HST} 
magnitudes denoted by $^*$). The optical photometry and recessional velocity for the Sombrero UCD 
is from \cite{Hau09}. All magnitudes have been corrected for foreground dust extinction following 
\citet{Schlegel1998}. K$_{\rm s}$ data for NGC4546 and the Sombrero are from 2MASS; K$_{\rm s}$ 
data for NGC3923 are from our reduction of archival NTT imaging. Recessional velocities are from 
SOAR/Goodman except for the Sombrero UCD.}
\label{tab:ucd_photometry}
\end{table*}

\subsubsection{NGC3923 GC Photometry}
\label{Sec:ngc3923gcs}

We have reexamined the GC system of NGC3923 in order to provide a sample of GCs to compare 
to our NGC3923 UCDs. The data used are the same \textit{HST} ACS data described above, which 
have been used previously to study the GC system of NGC3923 by \cite{Sikkema06} and \cite{ChoThesis}.

Our reanalysis of the data followed the procedure of \cite{ChoThesis}; we ran \textsc{SExtractor} on 
the \textsc{MultiDrizzle} reduced frames with the selection parameters set to a minimum area of 5 
pixels and a significance of at least 3$\sigma$ above the background computed in a 32$\times$32 
pixel box. The flux within a 3 pixel radius around each of the detected objects was then measured 
and an aperture correction made to correct the photometry to an infinite aperture (0.321~mag in 
F606W and 0.367~mag in F814W). As in the UCD case the photometry was corrected for Galactic 
extinction and the measured instrumental magnitudes were converted to V and I.

The major difference between our reanalysis and the earlier works is in the way potential GCs/UCDs 
are selected. Unlike \cite{Sikkema06} and \cite{ChoThesis} we do not require that all objects must be 
close to unresolved to be classed as GCs, nor do we apply a simple upper magnitude limit for inclusion 
in the sample. This is in marked contrast to studies of extragalactic GCs in general, where an upper 
magnitude limit close to that of $\omega$Cen is often applied to reduce contamination by Milky Way 
stars. It is our belief that this approach has most likely led to the misclassification of many potential 
UCDs as background galaxies or foreground stars in photometric surveys of extragalactic GCs.

Accepting partially resolved GC candidates is supported by the fact that five spectroscopically 
confirmed GCs \citep[][2011 in prep.]{Norris08} are significantly extended, with \textsc{SExtractor} 
\textsc{class}$\_$\textsc{star} values of 0.2 to 0.75, well below the limit imposed by \cite{ChoThesis} 
of 0.9. However, we do limit the GC analysis to objects fainter than M$_{\rm V}$ $>$ --11.0. Objects 
brighter than this must be resolved to be included in the UCD analysis. M$_{\rm V}$ = --11.0 is where 
the UCD luminosity-size trend ensures that bona-fide UCDs are large enough to be resolved at the 
distance of NGC3923 (see Section \ref{sec:Structural_Props}). 

To reduce contamination by stars and background galaxies we restricted our GC/UCD sample to 
objects with ellipticities less than 0.25, and imposed a colour cut of 0.7 $<$ V--I $<$ 1.5. This 
selection recovered all spectroscopically confirmed GCs, as well as the three NGC3923 UCDs. 
We restrict further analysis to objects with I $< $ 23.8 (0.5 mag brighter than the 90$\%$ 
completeness limit, see Appendix B), which ensures that we are dealing with the regime where 
incompleteness is negligible in both V and I. In total we find 549 objects that meet our criteria and 
that we provisionally classify as GCs in addition to the three UCDs.

\subsubsection{UCD Structural Modelling and Half Light Radii}
\label{sec:half_light}

We used $\textsc{galfit}$ \citep{Peng02_Galfit} to fit King and S\'{e}rsic models to 5$''$$\times$5$''$ 
(515$\times$515 pc) images of the NGC3923 UCDs. Background and foreground objects and bad 
pixels were masked, and position dependent artificial PSFs were constructed using $\textsc{TinyTim}$ 
\citep{tinytim}. Each PSF was 4$''$$\times$4$''$ in size and was produced using a G2V template.
We remove sky and galaxy background as described in Section \ref{Sec:HST_IP}. We also 
experimented with different median smoothing box sizes, finding that the resulting \textsc{galfit} model 
is not significantly affected as long as the box size is several times the half-light radii of the objects 
under study.

We fit both standard \citep{King62} and generalised King profiles \citep{Elson99}, which typically 
fit GCs well. We also fit S\'{e}rsic \citep{Sersic68} and generalised King + S\'{e}rsic models. In 
practice the S\'{e}rsic component mostly helps to fit unresolved core features. Note that in all 
cases the cores are unresolved (or the objects lack cores), so derived parameters such as central 
surface brightness are unreliable. However, properties that depend on large-scale modelling such 
as half-light radii are relatively robust. These radii underlie our analysis in Section \ref{sec:Structural_Props}.

To determine the intrinsic half-light radii R$_{\rm e}$ of our UCDs in the absence of PSF effects 
we used \textsc{galfit} to reconstruct  the best fit King + S\'{e}rsic models without the PSF convolution.
The half-light radii were then determined by direct integration of the model profiles. We quote the 
average of the radii determined from the F606W and F814W images for the NGC3923 UCDs. For 
the NGC4546 UCD only the F606W image is available. For the Sombrero UCD we quote the half-light 
radius determined by \cite{Hau09}, but find that our half-light radius for this UCD is consistent with 
theirs within the quoted errors (12.9 $\pm$ 2.0 pc compared to 14.7 $\pm$ 1.4 pc).
 
We are unable to place any strong constraints on the presence of low surface brightness envelopes 
around our UCDs. In the case of NGC4546 this is because the photometry is too shallow. For the 
NGC3923 UCDs it is due to the strong shells of NGC3923 which complicate accurate galaxy 
background removal.

\subsection{Stellar Mass Estimates}
\label{sec:stellar_mass_estimates}

To produce stellar mass estimates for the UCDs we use a modified version of the stellar mass 
estimation code first presented in \cite{Kannappan07} and later updated in \citet{KGB09}. Briefly, 
the code fits photometry from the Johnson-Cousins, Sloan, and 2MASS systems with an extensive 
grid of models from \citet{BruzualCharlot} (the updated 2009 models which include the effects of 
asymptotic giant branch stars) assuming a Salpeter initial mass function (IMF). The derived masses 
are then rescaled by a factor of 0.7 to approximate the ``diet" Salpeter IMF of \citet{Bell01}. Each 
collection of input UCD photometry is fit by a grid of model SSPs with ages from 25~Myr to 13.5~Gyr 
and metallicity Z = 0.0001 to 0.05. Unlike \citet{KGB09} our input models are pure SSPs, with no 
composite models, and no extinction. The derived stellar mass is determined by the median and 
68$\%$ confidence interval of the mass likelihood distribution binned over the grid of models. We 
also measure stellar masses for the host galaxies using catalog Hyperleda + 2MASS photometry 
and exactly the same grid of models as in \citet{KGB09} to facilitate comparison of galaxy mass 
scales (Section \ref{sec:Individual_Cases}).

The estimation of stellar masses is inherently uncertain. In fact, the use of different estimation 
methods or SSP models can lead to changes in the derived masses of up to factors of two, even 
when assuming the same IMF \citep{Kannappan07}. This problem is compounded by the fact that 
the correct choice of IMF for UCDs is still a matter of debate \citep[see e.g. ][]{Mieske08_AN,Chilingarian10}. 
We therefore caution that UCD stellar mass estimates are subject to systematic uncertainties in 
overall normalization at a factor of $\sim$3 level, which should not strongly affect comparisons 
within a given sample analyzed uniformly.

\section{Results}
\label{Sec:results}

In this section we present several basic observational results, which are discussed in a common 
framework in Section \ref{sec:Discussion}. We examine the orbital characteristics, stellar populations, 
and statistical distributions of our UCDs to demonstrate that (i) the candidate UCDs are bona-fide 
and associated with their apparent hosts, (ii) their ages, statistics in relation to the GC luminosity 
function, and apparent decay timescales are distinct, pointing to different formation mechanisms 
for the NGC3923 and NGC4546 UCDs, and (iii) our data provide no evidence yet for dependence 
of UCD frequency on galaxy environment. We argue that UCDs that form as giant globular clusters
(GGCs), as in NGC3923, emerge with distinct structural properties above a mass of 2$\times$10$^6$ 
M$_\odot$ and likely cannot exceed a mass of 7$\times$10$^7$ M$_\odot$, while UCDs that form 
as stripped nuclei, as in NGC4546, can exist at all masses.

\subsection{UCD Recessional Velocities $\&$ Survival Times}
\label{sec:UCD_RV_DFDT}

As detailed below, recessional velocities derived from SOAR spectroscopy confirm physical 
association between the UCDs and their host galaxies. Examination of the velocities and projected 
radii of the UCDs indicates a possible difference in dynamical friction decay timescale between the 
NGC4546 UCD and the other UCDs.

\begin{figure} 
   \centering
   \begin{turn}{0}
   \includegraphics[scale=1.0]{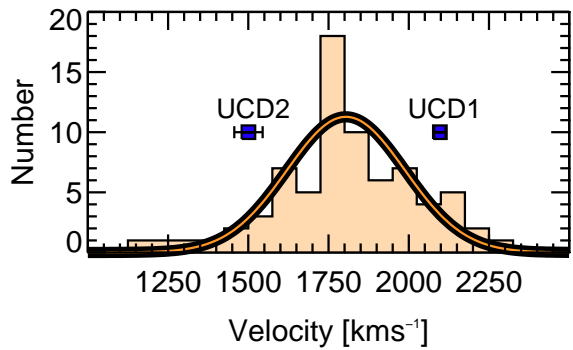}
    \end{turn} 
   \caption{Histogram of NGC3923 GC velocities measured by Norris et al. (2008, 2011 in prep).
   Overplotted are the best-fit Gaussian to the GC population and the velocities of the two confirmed 
   NGC3923 UCDs (squares).}
   \label{fig:ngc3923_gcs_ucds_vel}
\end{figure}

In the case of NGC3923 the GC velocity distribution has been measured by \citet{Norris08} and 
will be studied in detail in a forthcoming paper (Norris et al. 2011, in prep.). As 
Figure~\ref{fig:ngc3923_gcs_ucds_vel} shows, the velocities of NGC3923-UCD1 and UCD2 
(2096.9$\pm$17.8, 1500.5$\pm$44.2~kms$^{-1}$) are in excellent agreement with those of the GC 
system of NGC3923 (as well as NGC3923 itself, which has V=1739~kms$^{-1}$) confirming physical 
association. In further analyses we assume that UCD3c is also physically associated with NGC3923. 
The recessional velocity of NGC4546 is 1050 $\pm$ 9~kms$^{-1}$, therefore our measured recessional 
velocity for NGC4546-UCD1 of 1255.9$\pm$24.0~kms$^{-1}$ implies a high likelihood that these two 
objects are physically associated. Finally, the UCD of the Sombrero has a relative velocity 
(269$\pm$11~kms$^{-1}$) that is slightly larger than the one standard deviation of the GC velocity 
distribution of the Sombrero, making it likely that this UCD is bound to the Sombrero galaxy \citep{Hau09}.

Making the simplifying assumption of circular orbits, and assuming that the projected separations of 
the UCDs from their host galaxies are of the same order of magnitude as the true physical separations 
r$_{\rm i}$, we expect their survival times to obey the dynamical friction timescale from equations 8.2 
and 8.12 of \citet{BinneyTremaine08}, applicable for small satellites of massive galaxies:

\begin{eqnarray}
t_{\rm fric} &=& \frac{19~\rm Gyr}{{\rm ln}{\Lambda}} \left(\frac{r_{\rm i}}{5~{\rm kpc}} \right)^2 \frac{\sigma}{200~{\rm kms^{-1}}}\frac{10^8~{\rm M_\odot}}{M}.
\end{eqnarray}

\noindent Here $\sigma$ is the velocity dispersion of the host galaxy, \textit{M} is the total mass 
of the UCD (as determined in Section \ref{sec:ucd_masses}), and

\begin{eqnarray}
\rm ln{\Lambda} &=& {\rm ln} \left(\frac{b_{\rm max}}{r_{\rm h}} \right).
\end{eqnarray}

\noindent We assume that r$_{\rm h}$, the half-mass radius of the UCD, is equal to r$_{\rm e}$, 
the half-light radius of the UCD, and that b$_{\rm max}$ can be approximated by the observed 
projected separation between the UCD and host galaxy centre.

The resulting dynamical friction decay timescales for our UCDs are provided in Table~\ref{tab:ucd_structural_properties}. It is intriguing that the UCDs of NGC3923 and the Sombrero have 
long dynamical friction decay timescales of at least 7~Gyr, and usually several Hubble times,
whereas the UCD of NGC4546 is only expected to survive for around 0.5~Gyr. Objects that are 
already long-lived at the present would not be expected to have short dynamical friction decay 
timescales on average, as only a handful should be caught fortuitously at the end of their lives. 
This implication is borne out by the fact that none of the the GC candidates (modulo the ones lost 
due to the bright background in the very centre) detected by the HST have computed decay 
timescales of less than 8~Gyr. Moreover, based on galaxy and UCD/GC masses alone, the 
NGC3923 UCDs have normal dynamical friction decay timescales, i.e the UCDs of NGC3923 
appear to be drawn from the same population as the GCs of NGC3923. In contrast, after correcting 
for differences in host galaxy mass, the NGC4546 UCD is consistent with outliers comprising only 
1.8$\%$ of the distribution of cluster mass vs. decay time. Nonetheless, with only a handful of 
UCDs we cannot be sure that the projected quantities reflect the true values.

\subsection{UCD Stellar Populations}

In this section we examine the stellar populations of our UCDs. In the case of the NGC4546 UCD 
we utilise SAURON spectroscopy to measure its age, metallicity, and alpha-element enhancement, 
which differ significantly from those of the host galaxy. For the Sombrero, the stellar population 
information provided by \cite{Hau09} and \cite{Larsen02} indicates similar UCD and GC properties, 
with both having old ages (age $>$ 10 Gyr) consistent with those measured for the centre of the 
Sombrero by \cite{SanchezBlazquez06b}. Unfortunately we currently lack spectroscopy of sufficient 
S/N to examine the stellar populations of the NGC3923 UCDs. However, the NGC3923 GCs, 
including several which are marginally resolved by \textit{HST} and hence may prove to be UCDs, 
have been studied by \citet{Norris08}, with all studied GCs consistent with being old. Furthermore, 
a general examination of the position of UCDs, GCs, and dwarf nuclei in colour-magnitude space 
reveals that all three populations significantly overlap. In particular, blue UCDs and dwarf nuclei 
extend the observed ``blue tilt" or mass-metallicity relation of blue GCs.

\subsubsection{The Young UCD of NGC4546}
\label{sec:ngc4546stellarpops}

We find that the single UCD of NGC4546 is considerably younger than its host galaxy 
(3.4 vs. 10.7~Gyr), with significantly reduced alpha-element enhancement ($\sim$0 vs. 0.3), implying 
extended star formation.

\begin{figure} 
   \centering
   \begin{turn}{0}
   \includegraphics[scale=0.75]{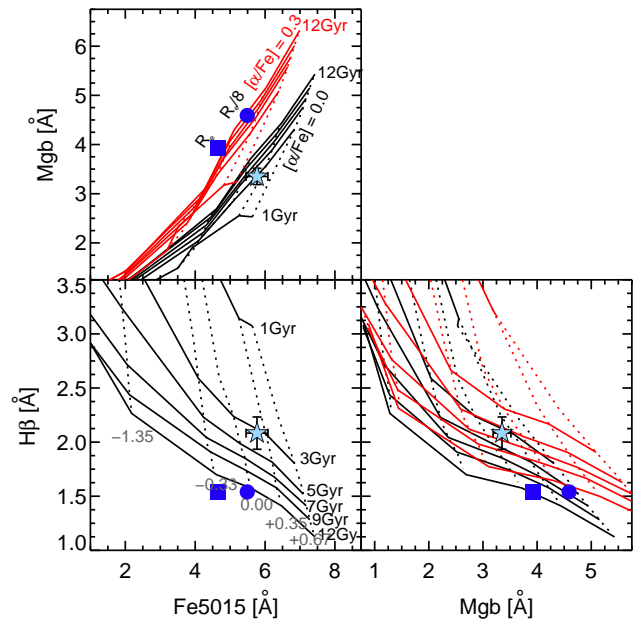}
   \end{turn} 
   \caption{Index-index plots for the NGC4546 UCD (blue star), measured as described in the text, 
   plus the diffuse light of NGC4546 measured at R$_{\rm e}$/8 (blue circle) and R$_{\rm e}$ 
   (blue square) by \citet[][]{Kuntschner06}. Overplotted are the model grids from \citet{Thomas03,Thomas04} 
   spanning the age range of 1 to 12~Gyr and [Z/H] from --2.25 to +0.67. For index-index combinations 
   with sensitivity to [$\alpha$/Fe] variation we overplot two choices of [$\alpha$/Fe], appropriate for the 
   UCD ([$\alpha$/Fe] = 0.0, black grid) and the NGC4546 diffuse light ([$\alpha$/Fe]=0.3, red grid).}
   \label{fig:SAURON_Lick}
\end{figure}

Figure \ref{fig:SAURON_Lick} displays the Lick/IDS absorption line indices determined for the UCD 
in Section \ref{sec:Sauron_spec}, as well as values measured for the diffuse light of NGC4546 at 
R$_{\rm e}$/8 and R$_{\rm e}$ by \citet{Kuntschner06,Kuntschner10}. Overplotted are the SSP 
models of \citet{Thomas03,Thomas04}. For index combinations sensitive to the effects of [$\alpha$/Fe] 
(those with Mg$b$) we overplot two choices of [$\alpha$/Fe], the black grid has [$\alpha$/Fe]=0.0, 
appropriate for the UCD, the red grid has [$\alpha$/Fe]=0.3, appropriate for the NGC4546 diffuse 
light. As Figure~\ref{fig:SAURON_Lick} demonstrates the implied stellar population parameters of 
age, [Z/H] and [$\alpha$/Fe] are consistent for all choices of index combination. By eye the UCD 
appears to be relatively young ($\sim$3~Gyr), metal rich ([Z/H] $\sim$ solar) and with [$\alpha$/Fe] 
close to solar. In contrast, the diffuse light of NGC4546 is old, displays a negative metallicity gradient, 
and is alpha-enhanced.

In order to determine a more definitive best-fit SSP we use the $\chi$$^2$-minimisation approach 
of \citet{Proctor04}. To carry out the fit we first interpolate the SSP models of \cite{Thomas03,Thomas04} 
to a finer model grid, then we perform a $\chi$$^2$-minimisation on the resulting age, [Z/H] and 
[$\alpha$/Fe] space. Errors come from 50 Monte-Carlo resimulations of the input data within the 
measured index errors. The final best-fit SSP model for the NGC4546 UCD is: 
age = 3.4$^{+1.7}_{-1.2}$~Gyr, [Z/H] = 0.21 $\pm$ 0.14, and [$\alpha$/Fe] = --0.01 $\pm$ 0.08. 
The equivalent SSP parameters for the diffuse light of NGC4546 as measured by \cite{Kuntschner10} 
at R$_{\rm e}$ are: age = 11.7$^{+1.1}_{-1.0}$~Gyr, [Z/H] = 0.13 $\pm$ 0.02, and [$\alpha$/Fe] = 0.31 
$\pm$ 0.05. At R$_{\rm e}$/8 the parameters are: age = 10.7$^{+1.0}_{-0.9}$~Gyr, [Z/H] = --0.13 $\pm$ 0.02, 
and [$\alpha$/Fe] = 0.27 $\pm$ 0.04. We discuss the implications of these results in Section 
\ref{sec:Discussion}.

\subsubsection{Indistinguishable Colour Magnitude Diagrams and Blue Tilts for GCs, UCDs, and Galaxy Nuclei}
\label{sec:blue_tilts}

The ``blue tilt" is the observation that in massive galaxies the more luminous blue population 
GCs are systematically redder (more metal rich) than their lower luminosity counterparts 
\citep[e.g.][]{Strader06,Mieske06,Harris06,Peng09,Forbes10b}. In this section we demonstrate 
that blue GCs, blue UCDs, and dwarf nuclei display the same blue tilt in their colour-magnitude 
relations. Furthermore, we observe that the nuclei of dwarf galaxies display identical colour-magnitude 
behaviour to that of GCs at lower mass, and UCDs at higher mass (including following the same 
steepening of the tilt above $\sim$M$_{\rm V}$=--10). Assuming that the observation that GCs, 
UCDs and dwarf nuclei display indistinguishable blue tilts is not a conspiracy of the age-metallicity 
degeneracy and that all three types of object have similar ages, this striking correspondence can 
be interpreted in three subtly different ways. 1) Dwarf nuclei are merely GCs/UCDs that have sunk 
to the centre of their host galaxies. 2) Blue GCs/UCDs are the stripped nuclei of dwarfs. 3) The 
same physics of self enrichment acting on all three object types leads independently to the same 
mass-metallicity relation. The implications of these explanations will be discussed in detail in 
Section \ref{sec:Discussion}.

\begin{figure*} 
   \centering
   \begin{turn}{0}
   \includegraphics[scale=0.8]{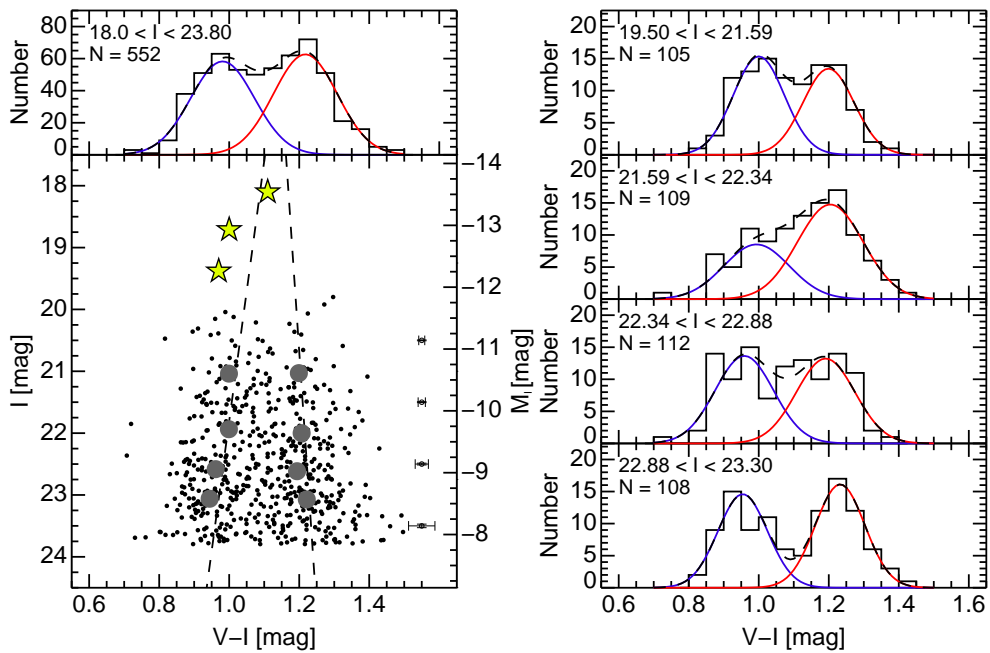}
   \end{turn} 
   \caption{$\bf{Left\,Panel:}$ Lower: V--I colour-magnitude diagram of 549 NGC3923 GCs (small 
   black dots) and UCDs/UCD candidates (yellow stars). Large filled circles denote the positions of 
   the peaks of the best-fit KMM \citep{KMM} two Gaussian models to the GC colour distributions in 
   four magnitude bins. The dashed lines denote the best-fit linear relations to the large filled circles. 
   Upper: The upper panel shows the total colour distribution of all 552 objects (black histogram), 
   and best-fit blue and red GC colour distributions and their sum (blue, red and black lines).
   $\bf{Right\,Panel:}$ Same as in upper left panel but for the four magnitude bins into which the GCs 
   have been divided for the examination of the blue tilt (see Section \ref{sec:blue_tilts} and 
   Appendix \ref{sec:Appendix_Blue_Tilt}).}
   \label{fig:ngc3923_cm}
\end{figure*}

Figure \ref{fig:ngc3923_cm} displays the NGC3923 GC/UCD colour-magnitude diagram. Bimodality 
in the colours of the GC system is detected with high significance \citep[$>$99$\%$, as estimated using 
the KMM method of][]{AshmanBirdZepf}. The measured peaks of the blue and red GC populations of 
V--I = 0.97 and 1.23 are close to those found for the GC systems of ellipticals such as M87 
\citep[V--I = 1.0 $\&$ 1.2,][]{Tamura06}. They are also consistent with those determined previously for 
the NGC3923 GC system by \citet{ChoThesis}.

Intrigued by the possible existence of a ``blue tilt" in the blue GCs of NGC3923 we estimated the 
significance of such a trend. A description of the procedure used is provided in Appendix 
\ref{sec:Appendix_Blue_Tilt}. The tilts measured in this investigation can be seen in Figure 
\ref{fig:ngc3923_cm} as the dashed lines. A blue tilt in the blue GC population is detected at 
$>$ 3$\sigma$ confidence, with a slope indistinguishable from that displayed by the GCs around 
M87 \citep{Peng09}. No evidence is found for a corresponding tilt in the red GC system.

\begin{figure*} 
   \centering
   \begin{turn}{90}
   \includegraphics[scale=0.90]{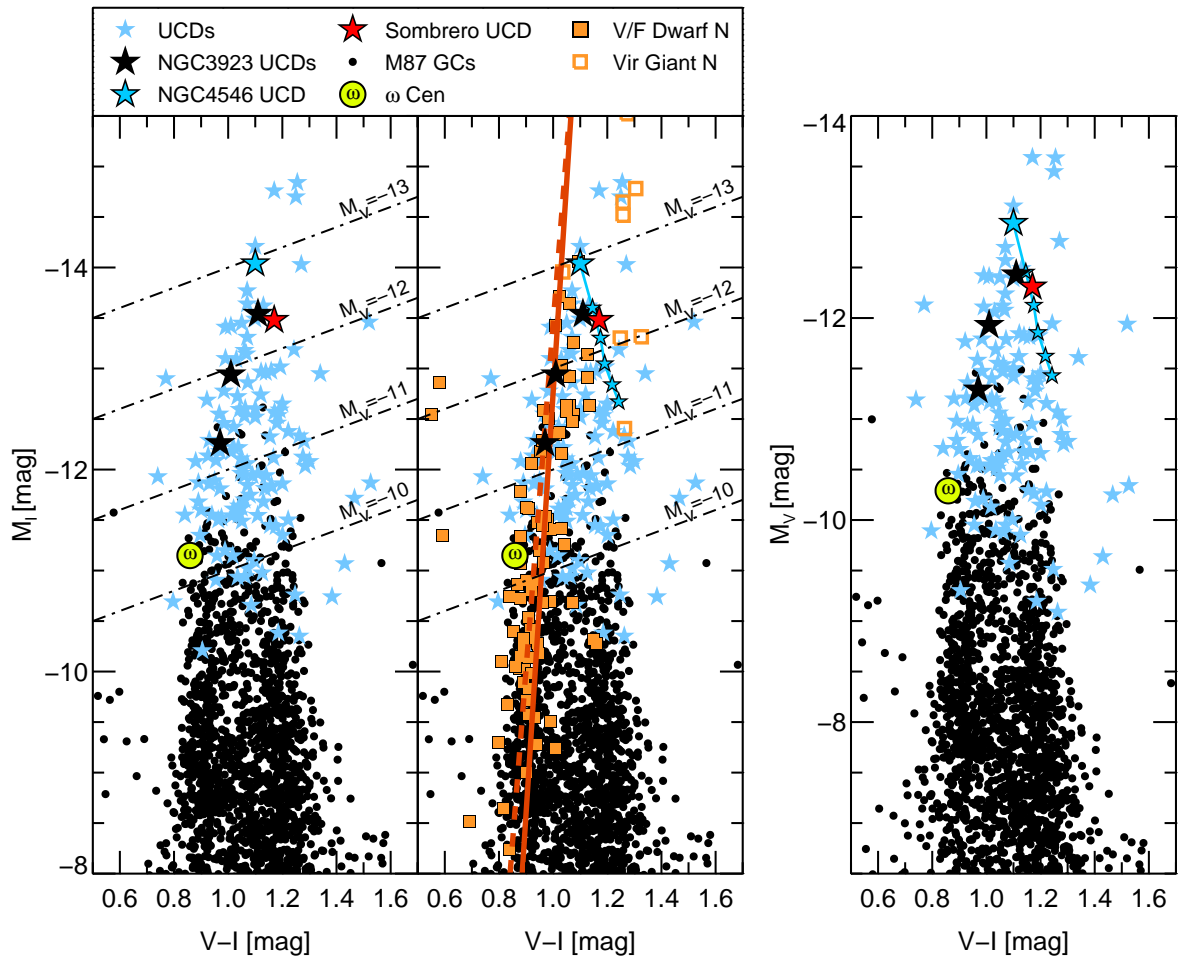}
   \end{turn} 
   \caption{Colour-Magnitude Diagrams for M87 GCs (black dots), UCDs (stars), and Virgo and 
   Fornax cluster galaxy nuclei (squares). The solid black stars denote our NGC3923 UCDs. The 
   large blue star is the predicted position of the NGC4546 UCD based on its spectroscopically 
   measured stellar population (see Section \ref{sec:ngc4546stellarpops}), while the small blue 
   stars are predictions for the evolution of the NGC4546 UCD for ages of 5, 7, 9, 11 and 13~Gyr. 
   The red star is the predicted position of the Sombrero UCD based on the measured SSP from
   \citet{Hau09}. Orange squares represent Virgo, Fornax and Leo group dwarf nuclei from 
   \citet{Cote06} and \citet{Lotz04}, with filled/open squares representing nuclei of galaxies with 
   M$_{\rm B}$ fainter/brighter than --17.6 respectively. The M87 GC photometry is from \citet{Peng09}, 
   and the UCD photometry is from the catalog of UCDs described in the text. The solid line is the 
   best-fit ``blue tilt" relation found for M87 blue GCs by \citet{Peng09}. The dashed line is the 
   best-fit relation found for the nuclei of galaxies with M$_{\rm B}$ fainter than --17.6 (identical 
   within the errors).}
   \label{fig:M87_GCs_UCDs}
\end{figure*}

Of particular interest in Figure \ref{fig:ngc3923_cm} is the fact that the NGC3923 UCDs are all 
consistent with an extrapolation of the measured blue tilt (all three are within 1.5$\sigma$ of the
best fit trend). This perhaps indicates a continuation 
of this phenomenon into the UCD mass range. In order to investigate this behaviour further we 
have constructed a catalog of published UCD photometry and compared this to the behaviour of 
the M87 GCs observed by \citet{Peng09};  see Figure \ref{fig:M87_GCs_UCDs}. We choose M87 
as our comparison sample because of the very high quality V and I band photometric measurements 
of 2250 GCs provided by \citet{Peng09}. We note that the measured value of the blue tilt varies from 
galaxy to galaxy with a dependence on galaxy mass \citep{Mieske10}. However, almost all of the 
UCDs to be discussed here (except the NGC4546 UCD) are likely associated with high-mass 
galaxies (those defined by \citealt{Mieske10} as having M$_*$ $>$ 5$\times$10$^{10}$~M$_\odot$), 
which \citet{Mieske10} find to have the largest magnitude blue tilts. Therefore, the scatter added 
due to the possible superposition of several varying UCD blue tilts should be relatively small.

The UCD photometry compiled for Figure \ref{fig:M87_GCs_UCDs} consists of our field/group 
UCDs plus Virgo \citep{Hasegan05,Firth08,Evstigneeva08}, Fornax \citep{Mieske04,Firth07,Firth08,Evstigneeva08}, Centaurus \citep{Mieske09} and Hydra \citep{Misgeld08} cluster 
UCDs/DGTOS/CSSs\footnote[2]{In the future we will refer to all UCDs/DGTOs/CSSs as UCDs.}  
observed directly in V and I or in equivalent bands (g/r/i or g/z) converted into V and I (using the 
SDSS to Johnson-Cousins conversions of Lupton\footnote[3]{www.sdss.org/dr7/algorithms/sdssUBVRITransform.html},
or the conversions of \citealt{Evstigneeva08}, assuming the UCDs are 11--13~Gyr old). To check 
the accuracy of the conversions to V/I, we compare native V/I magnitudes to those derived from 
the conversions for a sample of $\sim$20 UCDs that have both V/I and g/r/i or g/z observations. 
We find small $\delta$V, $\delta$I $<$ 0.1~mag systematic offsets, which we subtract off all converted 
V/I magnitudes to ensure consistent photometry. The dispersion in the native V/I vs. converted V/I 
photometry is around 0.1~mag, making this source of error comparable to that of the standard 
photometric errors. To place the NGC4546 and Sombrero UCDs (that lack I band photometry) on 
the same figure we adopt a different approach. We predict V and I fluxes from the measured SSP 
ages and metallicities found in Section \ref{sec:ngc4546stellarpops} for NGC4546 and by \citet{Hau09} 
for the Sombrero UCD, making use of the SSP models of \citet{Maraston05}. This method reproduces 
colours within 0.1~mag for all of the optical colour combinations we have access to (B--V, V--R 
and B--R).

On Figure \ref{fig:M87_GCs_UCDs} we also overplot a sample of dwarf nuclei. The dwarf nucleus 
measurements are provided by \cite{Cote06} for Virgo dwarfs (converted from g--z using the 
approach described above) and \cite{Lotz04} for a sample of Virgo, Fornax and Leo group dwarfs
(observed in V/I).

Focusing on the left panel first, of immediate note is the fact that there is considerable overlap 
in magnitude between objects classified as GCs and those classified as UCDs. This is likely 
due to the fact that the \textit{HST} imaging necessary to securely distinguish UCDs and GCs 
on the basis of size does not exist for all objects; many of these lower luminosity UCDs are in
all probability bright intra-cluster GCs.

Also of interest is the observation that the UCDs seem to continue the colour-magnitude trends 
of GCs. The vast majority of UCDs lie within the lower and upper colour limits of the GCs. The 
bluer UCDs, as seen for NGC3923 in Figure \ref{fig:ngc3923_cm}, also display a blue tilt similar 
to that of the blue GCs, including a steepening of the relation for M$_{\rm V}$ brighter than --10. 
This behaviour is strikingly reminiscent of that seen for the most luminous blue GCs of eight 
brightest cluster galaxies studied by \citet{Harris06}. It is also in reasonable agreement with the 
self-enrichment model of GC formation of \citet{Bailin_Harris09} (see their Figure 7). In the Bailin 
$\&$ Harris model the change in slope is a manifestation of the fact that clusters with mass $>$ 
10$^6$ ~M$_\odot$ (M$_{\rm V}$ $\sim$ --9) retain a significant amount of supernovae ejecta, 
leading to substantial self-enrichment.

Turning to the middle panel of Figure \ref{fig:M87_GCs_UCDs}, we note that dwarf nuclei trace 
the behaviour of the blue GCs and UCDs over six magnitudes in I, even displaying a similar 
break in slope (at M$_{\rm I}\sim$-11) . When we fit a linear relation to the dwarf nuclei over the 
same magnitude range as used by \cite{Peng09} to determine M87's blue tilt (--12 $<$ M$_{\rm I}$ $<$ --7.8),
the resulting fit is statistically indistinguishable from that of the M87 blue GCs. Also intriguing is 
the difference in behaviour between the nuclei of dwarf and giant galaxies. The dwarf galaxy 
nuclei display a strong blue tilt, while the giant galaxy nuclei are redder with fairly constant 
colour.  Although the M$_{\rm B}$~=~--17.6 mag dwarf/giant separator used by \cite{Cote06} 
was originally a morphologically defined separation between dwarf and giant Virgo galaxies,
it also corresponds to the ``gas-richness threshold mass" scale that lies at the transition point 
of several galaxian properties \citep{KGB09}. Using Nearby Field Galaxy Survey Tully-Fisher 
relation data \citep{Kannappan02} we find that M$_{\rm B}$~=~--17.6 corresponds to a circular 
rotation velocity of V$_{\rm circ}$ $\cong$ 120~kms$^{-1}$ for the red/old galaxies expected in 
clusters. This mass scale is where \citet{Garnett02} and \citet{Dalcanton04} observe sharp 
changes in [Fe/H] and dust lane structure, apparently simultaneously with strong shifts in gas 
richness (\citealt{KW08}, updating \citealt{Kannappan04}) and in galaxy structure \citep[][and references therein]{KGB09}. The colour dichotomy of galaxy nuclei therefore appears to be another important 
transition in galaxian properties occurring at the threshold mass scale. Notably, the expected 
colour-magnitude evolution of the young NGC4546 UCD may move it into the regime of red 
nuclei in $\sim$4 Gyr (see Section \ref{sec:Individual_Cases}).

\subsection{UCD Structural Properties}
\label{sec:Structural_Props}

In this section we examine the structures of our UCDs. We find that they are reasonably fit by 
GC-like King profiles. We also find that the half-light radii of our UCDs, 12 -- 25~pc, are 
considerably larger than those of typical GCs. As seen in previous work, we find that UCDs 
display luminosity-size and mass-size trends, in the sense that more luminous/massive UCDs 
are more extended. In contrast, GCs have roughly constant size over $\ga$ 2 dex in luminosity/mass. 
Additionally, we find a striking and previously unreported correspondence between the luminosity 
at which the M$_{\rm V}$-size trend of UCDs commences and the luminosity at which the ``blue tilt" 
(i.e. M$_{\rm V}$-metallicity trend), becomes most pronounced, which we jointly label the ``scaling 
onset luminosity".

\subsubsection{UCD Profile Shapes}

\begin{figure} 
   \centering
   \begin{turn}{0}
   \includegraphics[scale=0.69]{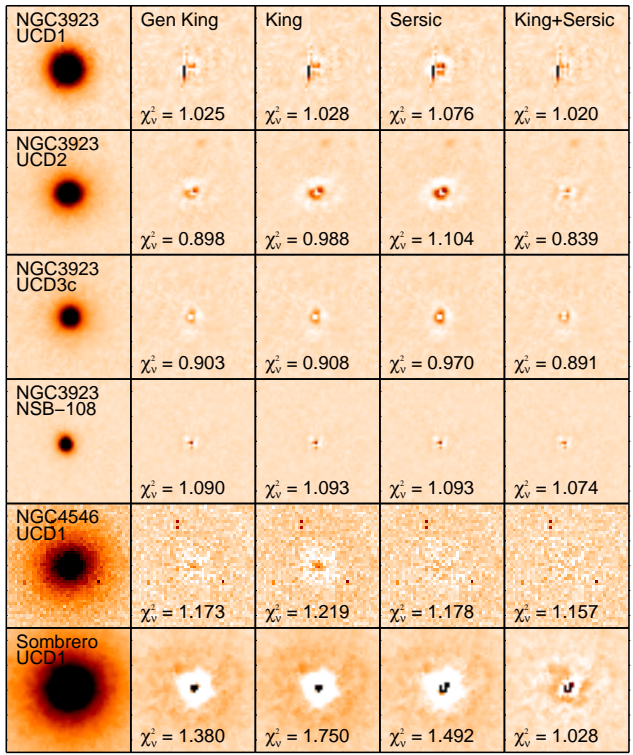}
   \end{turn} 
   \caption{Field/Group UCDs and NSB-108, one of the brightest spectroscopically confirmed 
   NGC3923 GCs. The first column displays the original F606W images (F625W for the Sombrero 
   UCD). The remaining columns show the residuals after subtracting the PSF-convolved GALFIT 
   models. All images are 2.5$^{\prime\prime}$$\times$2.5$^{\prime\prime}$.}
   \label{fig:ngc3923_ucds}
\end{figure}

Figure \ref{fig:ngc3923_ucds} shows result of fitting structural models to the UCDs and a bright 
(M$_{\rm V}$~=~--9.9) NGC3923 GC spectroscopically confirmed by \cite{Norris08}. As expected, 
the fit to the bright GC is unconstrained due to its marginal resolution. It is very difficult to draw firm
conclusions regarding the structures of the UCDs: they are all better fit by generalised King profiles 
or King+S\'{e}rsic profiles than by standard King or S\'{e}rsic models, but not significantly. There is 
no evidence for any particular UCD displaying a different structure from any other. In general, while 
the GALFIT modelling is useful in determining half-light radii for the UCDs (see Section \ref{sec:half_light}), 
the structures of our UCDs do not provide any clues to possible different formation scenarios within 
the UCD population.

\subsubsection{UCD Half-Light Radii}

Table \ref{tab:ucd_structural_properties} presents the half-light radii of our UCDs. The NGC3923 
UCDs have almost constant half-light radii of around 13~pc, the Sombrero UCD also has a similar 
half-light radius of 14.7~pc \citep{Hau09}, and the NGC4546 UCD is more extended at around 
25~pc. All five UCDs are therefore significantly larger (as well as more luminous) than Milky Way 
GCs, which have half-light radii of around 3.2~pc \citep{Rejkuba07}. These half-light radii place 
these objects squarely within the range displayed by Virgo and Fornax  UCDs as measured by 
\citet{Evstigneeva08}.

\subsubsection{UCD Luminosity/Mass-Radius Relations}
\label{sec:ucd_masses}

\begin{figure*} 
   \centering
   \begin{turn}{0}
   \includegraphics[scale=0.9]{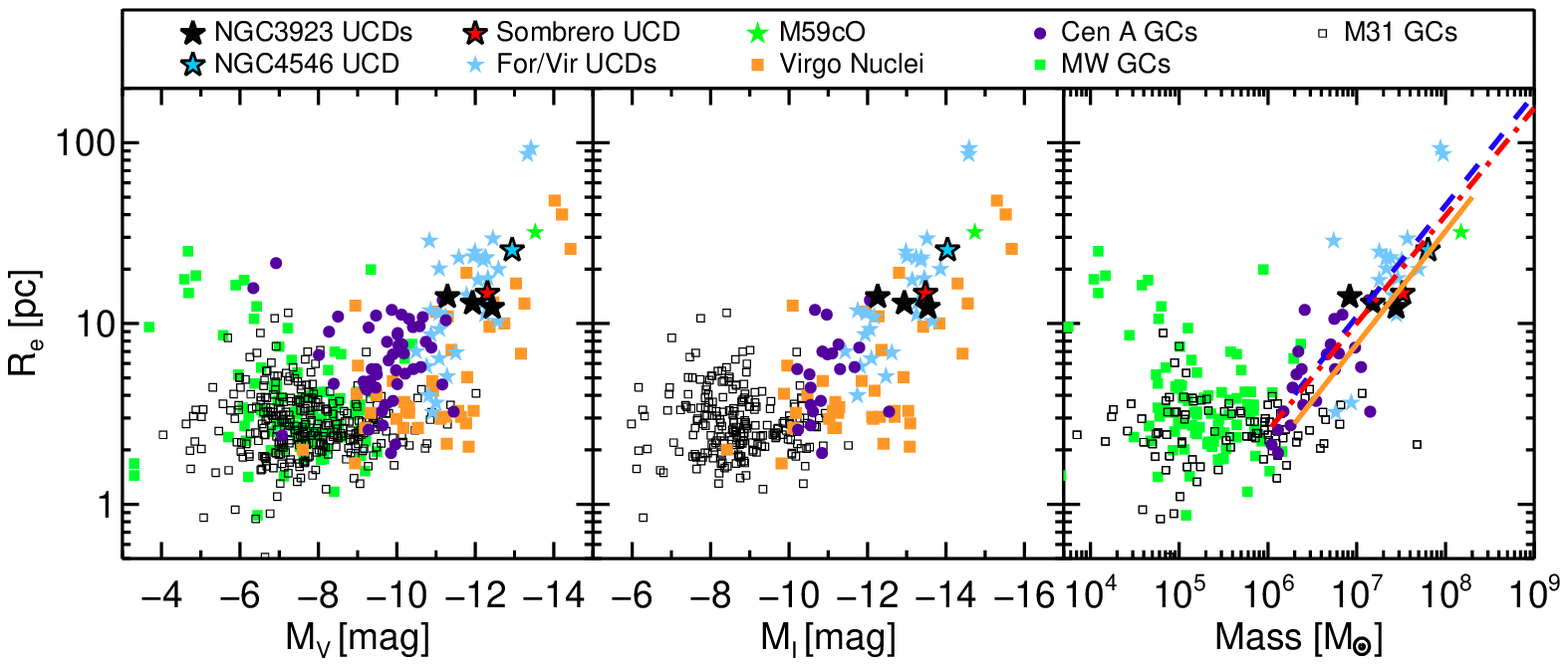}
   \end{turn} 
   \caption{$\bf{Left~and~Centre~Panels:}$ Half-light radius vs. V/I magnitude for compact stellar 
   systems. Milky Way GCs are from \citet{McLaughlin05}; M31 GCs are from \citet{Barmby07} 
   and \citet{Peacock09}; Cen A GCs are from \citet{Harris02}, \citet{Rejkuba07}, and \citet{Taylor10}. 
   Fornax and Virgo UCDs are from \citet{Evstigneeva07a,Evstigneeva08} and \citet{Hasegan05}.
   Large stars denote our NGC3923 UCDs, our NGC4546 UCD, and the Sombrero UCD from 
   \citet{Hau09}. The UCD M59cO is from \citet{Chilingarian08a}. Virgo dwarf nuclei are from 
   \citet{Cote06} converted to V as described in Section \ref{sec:blue_tilts}. $\bf{Right~Panel:}$ 
   Half-light radius against dynamical mass for GCs/UCDs. Sources as in left panel, except M31 
   GCs are from \citet{Barmby07} only and CenA GCs are from \citet{Taylor10} only. The masses 
   of our NGC3923 and NGC4546 UCDs are estimated dynamical masses, determined as described 
   in Section \ref{sec:ucd_masses}. The solid orange line is the model prediction of \citet{Murray09} 
   as described in Section \ref{sec:ucd_masses}. The dashed blue line is the extrapolation of the 
   relation for bright ellipticals from \citet{Hasegan05}, the red dot-dashed line is the best fit to bright 
   ellipticals, compact ellipticals, and galaxy bulges from \citet{Dabringhausen08}.}
   \label{fig:mass_size}
\end{figure*}

The left and centre panels of Figure \ref{fig:mass_size} demonstrate that our field/group UCDs fall 
along the same luminosity-size relation defined by cluster UCDs. To construct the equivalent 
mass-size relation seen in the right panel of Figure \ref{fig:mass_size}, we have estimated the 
dynamical masses of our UCDs assuming a constant dynamical to stellar mass ratio, calibrated 
on 18 UCDs that have both literature dynamical masses and stellar masses measured using the 
approach described in Section \ref{sec:stellar_mass_estimates}. From these 18 UCDs (spanning 
the dynamical mass range from 2$\times$10$^6$~M$_\odot$ to 1.5$\times$10$^8$~M$_\odot$) 
we find that the stellar mass of a UCD is 0.54$\pm$0.25 times the dynamical mass. With this 
correction we can estimate the total dynamical masses of our NGC3923 and NGC4546 UCDs. 
Using these masses, we find that our field/group UCDs conform to the same mass-size trends as 
cluster UCDs.

As previously noted for Milky Way and Virgo cluster GCs, most GCs display constant size with 
luminosity or mass \citep[see e.g.][]{Jordan05,Rejkuba07}. In contrast, UCDs exhibit increasing 
size with luminosity or mass \citep{Hasegan05,Evstigneeva08,Mieske08}. Intriguingly, the most 
luminous GCs (with M$_{\rm V}$ $<$ --10, or M $>$ 2$\times$10$^6$~M$_\odot$) of Cen A, the 
Milky Way ($\omega$Cen specifically) and M31 also seem to display a luminosity/mass-size 
trend. The nuclei of dwarf galaxies also display a mass-size trend, with similar slope to that of 
UCDs, though it appears to be offset to smaller radii (see Figure \ref{fig:mass_size} and 
\citealt{Evstigneeva08}). Like the similarities of these objects in the colour-magnitude diagram, 
these size trend similarities may be a sign of a comparable formation histories for GCs, UCDs and 
nuclei, or may merely be due to the influence of the same physical processes (e.g. mass loss by 
stellar evolution, evaporation and gravitational shocks).

If we use the onset of the luminosity/mass-size trend to classify objects as either GCs or UCDs 
we determine selection criteria very similar to those found in other studies \citep[e.g.][]{Mieske08}. 
UCD-like behaviour arises at around M$_{\rm V}$ = --10 (masses above~2$\times$10$^6$~M$_\odot$) 
and for objects more extended than $\sim$6 pc. There is however no clean switch in behaviour 
at any of these limits. This is clearly demonstrated by the existence of objects (M31 and Cen A 
GCs, Virgo dwarf nuclei, and Virgo UCDs) that have mass up to 10$^7$~M$_\odot$ whilst also 
having half-light radii entirely consistent with normal GCs. The existence of high-mass objects
that do not display a mass-size trend may indicate a second formation channel at work. Regardless, 
for any sensible GC/UCD separation, our field/group UCDs are all confirmed to be bona-fide UCDs. 

The onset of UCD-like behaviour at M$_{\rm V}$ $\sim$ --10 and R$_{\rm e}$ $>$ 6~pc, implies 
that several nominal GCs previously spectroscopically confirmed to be part of the NGC3923 
system by Norris et al. (2008, 2010 in prep.) potentially meet the UCD selection criteria. In fact, 
examination of the SExtractor CLASS$\_$STAR parameter for spectroscopically confirmed GCs 
shows that five ``GCs" are marginally resolved by the \textit{HST}. However, as the resolution limit 
of the ACS corresponds to a half-light radius of $\sim$10~pc at the distance of NGC3923, more 
sophisticated modelling is required to accurately measure the sizes of these objects. At present 
therefore, we have two confirmed UCDs around NGC3923, one high confidence candidate (UCD3c), 
and five other probable UCDs. Based on the analysis of spectra in \citet{Norris08} the stellar 
populations of these candidate UCDs match those of the exclusively old GCs of NGC3923.

We note with interest that the luminosities where UCD-like luminosity-size behaviour becomes 
apparent (see Figure \ref{fig:mass_size}) correspond well with the luminosities at which the blue 
tilt of GC systems is strongest \citep[compare Figs. \ref{fig:M87_GCs_UCDs} and \ref{fig:mass_size}; see also][]{Harris09}. This apparent concordance perhaps indicates that a common mechanism is responsible 
for the onset of both mass-size and mass-metallicity relations at  an equivalent ``scaling onset mass".

The cause of the UCD mass-size trend, and the reason for its sudden onset above 
$\sim$2$\times$10$^{6}$M$_\odot$ is still an open question, with several possible solutions, some 
of which are applicable to multiple of the potential UCD formation channels (single-cluster GGCs, 
merged GGCs, and stripped nuclei; see Section \ref{sec:intro}). 

\citet{Murray09} suggests that single star clusters of mass $>$ 10$^6$~M$_\odot$ are optically 
thick to far-IR radiation at formation, changing the behaviour of the Jeans mass systematically with 
cluster mass. This process leads to a top heavy IMF, which can also explain the observation that 
UCDs have higher mass-to-light ratios than GCs \citep{Chilingarian08b,Mieske08}. The predicted 
mass-size trend from \citet{Murray09} is overplotted on Figure \ref{fig:mass_size} as the solid line, 
providing a good qualitative match to the slope (if not the zeropoint) of the observed mass-size trend. 
The other predicted behaviours, such as the presence of a top-heavy IMF, and higher mass-to-light 
ratios provide a useful way to probe the validity of this scenario, and hence, the likely single collapse 
origin of some UCDs.

Complicating matters however, is the fact that a fit to the scaling relation for bright ellipticals (the 
dashed line in Figure \ref{fig:mass_size}, from Eqn 12. of \citealt{Hasegan05} or the darker dot-dashed 
line from \citealt{Dabringhausen08}) provides an even better fit to the UCD mass-size trend than the 
\citet{Murray09} prediction. This observation can itself have two explanations. The first is appropriate 
for all three possible UCD types (single-cluster GGCs, merged GGCs and dwarf nuclei): as suggested 
by \cite{Mieske08}, objects with the size and mass of UCDs (and larger) will be dynamically unrelaxed 
after a Hubble time, whereas GCs are sufficiently small that they have dynamically relaxed by the 
present. Recently \citet{Gieles10} explained this observation in terms of GCs evolving away from a 
common (with UCDs and elliptical galaxies) mass-size relation due to cluster expansion caused by 
hard binaries and mass loss in stellar evolution. This scenario does not in itself explain the existence 
of the mass-size trend, only the reason for GCs' deviation from the trend. Because of this, and because 
it is applicable to all potential UCD types, it does not provide any useful clues to the formation history 
of UCDs.

A second explanation for the close correspondence between the mass-size trends of UCDs and 
elliptical galaxies may be that UCDs form in a manner similar to elliptical galaxies, i.e. by violent 
relaxation after the merger of subcomponents. This explanation could be applicable to GGCs 
that form on short timescales via the merger of ``normal" GCs within a single giant molecular cloud. 
It may also apply to UCDs formed by the stripping of galaxy nuclei, because one possible formation 
scenario for galaxy nuclei is that they are built up by the merging of GCs that sink to the centre of 
their host galaxy by dynamical friction \citep[see e.g.][]{Agarwal10}. In this case, the observation
that galaxy nuclei are smaller at fixed mass than UCDs may imply that nuclei that are stripped to 
become UCDs must expand, e.g. as in \citet{Bekki03c}. Since in the merging scenario the existence 
of the mass-size trend for masses $>$ 2$\times$10$^6$~M$_\odot$ is directly related to the way 
in which the objects formed, this scenario is also consistent with the existence of objects with mass 
$>$ 2$\times$10$^6$~M$_\odot$ that do not display a mass-size trend, if they did not form by 
merging.

In each of these scenarios the cause of the common onset for the mass-size and mass-metallicity 
relations is mysterious. It is of course possible that this close agreement is merely a coincidence, 
however, any explanation that encompasses both behaviours is to be strongly preferred.

\begin{table}
\begin{center}
\begin{tabular}{lcccc} \hline
Name			&	Distance	&	R$_{\rm e}$			&	M$_\star$					&	t$_{decay}$\\
				&	[kpc]			&	[pc]				&	[$\times$10$^6$M$_{\odot}$]	&	[Gyr]	\\
\hline
\multicolumn{4}{l} {\textbf{NGC3923 UCDs}}\\
UCD1	&	3.23			&	12.3$\pm$0.3 (1.8)	&	15.0$^{+7.7}_{-7.1}$			&	7.3\\
UCD2	&	7.92			&	13.0$\pm$0.2 (3.2)	&	 8.2$^{+3.1}_{-3.7}$			&	71.5\\
UCD3c	&	6.14			&	14.1$\pm$0.2 (2.3)	&	 4.5$^{+0.9}_{-1.5}$			&	81.4\\
\multicolumn{4}{l} {\textbf{NGC4546 UCD}}\\
UCD1	&	1.16			&	25.5$\pm$(1.3)		&	34.3$^{+6.9}_{-10.6}$		&	0.5\\
\multicolumn{4}{l} {\textbf{Sombrero UCD}}\\
UCD1	&	7.50			&	14.7$\pm$1.4		&	33.0$^{+3.0}_{-3.0}$			&	25.2\\
\hline
\end{tabular}
\end{center}
\caption[UCDs]{UCD Structural Properties. Distance is the projected distance between the 
UCD and galaxy centre. Half-light radii are measured by direct integration of the best-fit GALFIT 
models as described in Section \ref{sec:half_light}. Uncertainties without parentheses are 
based on the measured differences between the half-light radii determined in the F606W 
and F814W bands; values in parentheses are uncertainties based on the difference between 
the PSF-convolved and unconvolved estimations. Properties for the Sombrero UCD are from 
\cite{Hau09}. Computation of UCD masses is described in Sections \ref{sec:stellar_mass_estimates} 
and \ref{sec:ucd_masses}. The calculation of dynamical friction decay time is described in
Section \ref{sec:UCD_RV_DFDT}.}
\label{tab:ucd_structural_properties}
\end{table}

\subsection{UCD Frequencies and Environments}
In this section we examine whether the numbers and luminosities of our UCDs can be statistically 
explained as the high luminosity tail of the globular cluster luminosity functions of their host 
galaxies, arguing that those around NGC3923 and the Sombrero can, while the one near 
NGC4546 cannot.  We also investigate the environmental dependence of UCD frequencies.

\subsubsection{GCLF Analysis}
\label{sec:gclf_analysis}

\citet{Hilker09} demonstrates that in most cases the absolute magnitude of the brightest GC/UCD 
of a galaxy correlates with the host galaxy luminosity and the total GC system size. Furthermore, 
he shows that the absolute magnitude of a galaxy of the brightest GC/UCD is well fit by a simple 
model where the observed GCLF is extrapolated to higher luminosities. This result implies that 
UCDs are in general the extreme high luminosity tail of the GC luminosity function. One natural 
result of such a scenario is that galaxies are often expected to host multiple UCDs, if their GC 
systems are large enough. The fact that we have discovered at least two, and as many as eight 
UCDs associated with NGC3923, would seem to lend support to their having been formed alongside 
the GCs of NGC3923. It is difficult to imagine that stripping of nuclei could produce so many similar, 
massive objects with long-lived orbits.

We wish to determine whether our field/group UCDs obey the \citet{Hilker09} correlations. Where 
possible we make use of literature determinations of the GCLF parameters; for the Sombrero we 
use the GCLF total size and dispersion measured by \citet{Rhode04}. Unfortunately, no suitable 
imaging data exists to allow an analysis of the GCLF of NGC4546. In the case of NGC3923 we use 
the determination of the total GC system size made by \citet{Sikkema06}, but unfortunately neither 
\citet{Sikkema06} nor \citet{ChoThesis} quote the dispersion of the GCLF of NGC3923 needed to 
predict the brightest GC/UCD. We have therefore examined the \textit{HST} photometry described 
in Section \ref{Sec:ngc3923gcs} to determine the dispersion of the NGC3923 GCLF. Full details of 
the procedure used to determine this are provided in Appendix \ref{sec:Appendix_GCLF}, see also 
Figure \ref{fig:NGC3923_GCLF}.

\begin{figure} 
   \centering
   \begin{turn}{0}
   \includegraphics[scale=1.00]{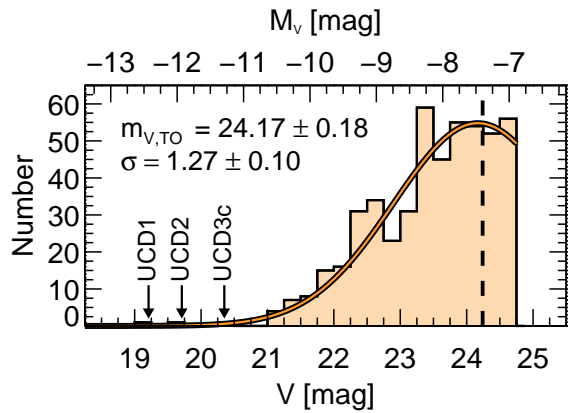}
   \end{turn} 
   \caption{NGC3923 V-band GC luminosity function. The best-fit Gaussian model is overplotted with 
   the best-fit parameters listed in the top left. Assuming the Table \ref{tab:galaxy_properties} distance 
   modulus of 31.64, the measured turnover magnitude equates to an absolute magnitude of M$_{\rm V}$ 
   = --7.47, consistent with the observed universal GCLF turnover magnitude of $\sim$~--7.4~$\pm$~0.1 
   shown by the dashed line \citep[e.g.][]{Kundu01a,Kundu01b,Jordan07}.}
   \label{fig:NGC3923_GCLF}
\end{figure}

\begin{figure*} 
   \centering
   \begin{turn}{0}
   \includegraphics[scale=0.90]{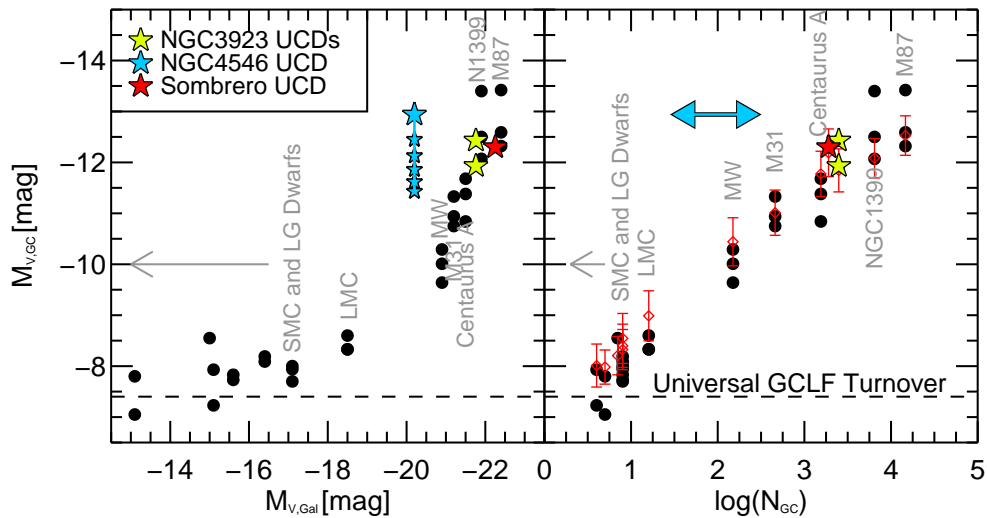}
   \end{turn} 
  \caption{Figure after \citet{Hilker09}. $\bf{Left~Panel:}$ The absolute magnitude of the brightest two 
  or three GCs/UCDs of a galaxy as a function of host galaxy total luminosity. Data come from the 
  compilation of \citet{Hilker09} plus estimates for NGC3923, NGC4546 and the Sombrero as described 
  Section \ref{sec:gclf_analysis}. The smaller blue stars are predictions for the future evolution of 
  NGC4546 UCD1 for ages (top to bottom) 5, 7, 9, 11 and 13~Gyr. $\bf{Right~Panel:}$ The absolute 
  magnitude of the brightest two or three GCs/UCDs of a galaxy as a function of the total number of 
  GCs present in the GC system of the host galaxy. The red diamonds with error bars indicate the 
  average luminosity of the brightest GC found in 10,000 Monte-Carlo simulations of the GCLF of each 
  galaxy, assuming the measured total number of clusters, the universal GCLF turnover and the measured 
  dispersion of the GCLF.  The blue arrowed region shows where NGC4546 UCD1 is likely to lie in this 
  plot.}
   \label{fig:simulations}
\end{figure*}

The left panel of Figure \ref{fig:simulations} shows the result of placing our field/group UCDs on the 
brightest GC/UCD vs. host galaxy luminosity plot. The UCDs of NGC3923 and the Sombrero are perfectly 
consistent with the overall trend of more luminous galaxies having more luminous brightest GC/UCDs. 
However, the NGC4546 UCD is substantially overluminous relative to this trend, being around 3 
magnitudes brighter than expected given the low luminosity of NGC4546. Even if we artificially evolve 
the NGC4546 UCD forward in age using the models of \cite{Maraston05} to ages 5, 7, 9, 11 and 13~Gyr 
(top to bottom small blue stars in Figure \ref{fig:simulations}), the NGC4546 UCD will still be significantly 
overluminous relative to its host galaxy.

The right panel of Figure \ref{fig:simulations} demonstrates that the magnitudes of the UCDs of NGC3923 
and the Sombrero correlate extremely well with total GC system size, again confirming Hilker's result.
Despite the fact that we cannot presently place the NGC4546 UCD on this plot, we can still state with 
some certainty that this UCD cannot be explained as bright extension of the GCLF. The horizontal arrow 
marks M$_{\rm V}$ for this UCD in the plot, showing that empirically it could occur in galaxies with GC 
systems as large as those in M87. However, for a galaxy of NGC4546's modest luminosity to produce 
such a large GC would require either the most unlikely statistical fluke, or that NGC4546 has a most 
unusual GC system, being extremely large for its host mass. Moreover, we suspect that the brightest 
UCDs around NGC1399 and M87 cannot be formed as GCs, based on statistical arguments. Further 
support for the idea that the brightest NGC1399 and M87 UCDs are in fact stripped nuclei comes from 
the presence around both of extended low surface brightness envelopes \citep{Evstigneeva08}, of the 
sort expected to persist after stripping.

The red diamonds in Figure \ref{fig:simulations} show predictions of a model in which two effects 
conspire to determine the expected magnitude of the brightest GC/UCD of a galaxy. The first is simply 
the increasing number of GCs found in larger galaxies. The second, more subtle effect, is that the 
dispersion of the GCLF of a galaxy depends on galaxy luminosity/mass \citep{Jordan07}, such that 
more luminous galaxies have broader GCLFs, making higher luminosity GCs more likely.

Our model estimates the \textit{expected} magnitude of the brightest GC/UCD of an observed GC 
system, as equal to the average magnitude of the brightest GC found in 10,000 Monte-Carlo simulations 
using the observed total GC system size, observed GCLF dispersion, and an assumed universal GCLF 
turnover magnitude of M$_{\rm V}$ = --7.4. As found by \citet{Hilker09} the predictions of this simple 
model are remarkably close to those found for galaxies from dwarfs to BCGs. The only exceptions appear 
to be the brightest UCDs of the NGC1399 and M87 systems. These objects are too luminous to be 
explained as an extension of the GCLF of NGC1399 or M87. For these UCDs to be luminous GCs would 
require that either: the total number of GCs present in the M87 and NGC1399 systems is currently 
underestimated by greater than a factor of 5, or, the dispersion of the GCLF of these galaxies is 
underestimated by $\sim$25$\%$. 

We consider errors of this size unlikely, and therefore suggest \textit{that M$_{\rm V}$ of around --13 
represents the practical upper magnitude limit for the existence of UCDs formed as GGCs.} GC 
systems of the required richness to produce larger UCDs aren't yet known, even relying only on 
statistical arguments and without bringing up physical objections to producing the large giant 
molecular clouds required. Above this limit UCDs are likely to be the result of other processes, 
such as the stripping of galaxy nuclei, while below this limit either process could be responsible, 
but GC-like formation seems most common, given the typically close correspondence between 
GC numbers and GC/UCD maximum luminosity.

\subsubsection{UCD Environmental Dependence}

\begin{figure*} 
   \centering
   \begin{turn}{0}
   \includegraphics[scale=1.0]{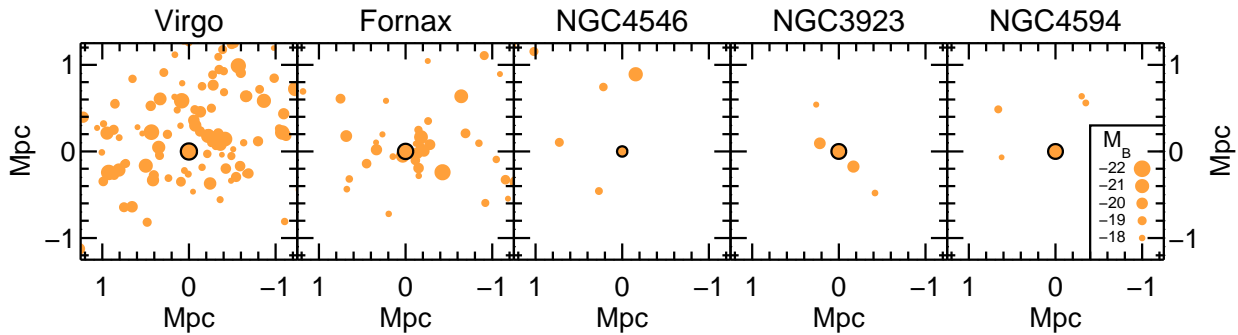}
   \end{turn} 
   \caption{
    Illustration of the relative environmental densities of the galaxies studied here. The equivalent 
    panels for the Virgo and Fornax clusters are provided to facilitate comparison with the denser 
    environments where UCDs are generally studied. In each panel filled dots indicate galaxies 
    found to be members of the same structure as the galaxy of interest by \citet{Giuricin00}, where 
    we apply a magnitude cut at M$_{\rm B}$ $<$ --17.65 (the magnitude limit at the distance of the 
    most distant group, NGC3923 and coincidentally similar to the dwarf-giant luminosity division 
    discussed in Section \ref{sec:blue_tilts}). The size of the filled dots is proportional to the M$_{\rm B}$
    of each galaxy. The central galaxies hosting UCDs are indicated by black circles (M87, NGC1399, 
    NGC3923, NGC4546 and the Sombrero respectively).}
   \label{fig:environment}
\end{figure*}

Our discovery of at least three further UCDs in non-cluster environments takes the number of 
UCDs found outside of high-density environments to five. Clearly UCDs are not simply a high 
density environment phenomenon. In fact, as Figure \ref{fig:environment} demonstrates, UCDs 
are found everywhere from galaxy clusters such as Virgo and Fornax, through groups such as 
the NGC3923 group, to loose group/isolated galaxies such as NGC4546 and the Sombrero. 
This observation does not itself rule out any of the UCD formation scenarios, as all possible 
formation routes are expected to occur in all environments, though with varying levels of efficiency.

Within the limits of present samples, it is not yet possible to draw firm conclusions about the 
environmental distribution of UCDs, as much larger and more carefully constructed samples 
making use of wider field HST imaging (for accurate GC/UCD separation) than available presently 
would be required for that. However, the number of UCDs found by our mini survey is in line with 
previous estimates of UCD frequency in clusters. Given that the ACS field of view covers an area 
of 0.003~deg$^2$ and that the WFPC2 covers 0.001~deg$^2$, we can use the distances to each 
of our 76 examined galaxies to estimate the area probed. In total we find that we have examined 
0.022~Mpc$^2$. If we make the assumption that UCDs are as common in all environments as in 
the Fornax cluster \citep[105$\pm$13 per circle of radius 0.9$^{\circ}$,][]{Gregg09}, then there 
should be 375~UCDs/Mpc$^2$. Therefore, given the area surveyed we would expect to find 9 
UCDs, a number in good agreement with our 11 candidates (4 spectroscopically confirmed). If 
we also consider that the number of UCDs found by \citet{Gregg09} undoubtedly includes some 
interloping intracluster GCs (as \citealt{Gregg09} lack the high resolution imaging required to 
measure radii), then even without any further UCD candidates being confirmed, the number of 
UCDs in field/group environments appears similar to the number in galaxy clusters.

\section{Discussion}
\label{sec:Discussion}

The preceding sections have presented multiple properties of a sample of confirmed field/group 
environment UCDs.  We now synthesise the implications of these observations for the formation 
of these UCDs in particular, as well as the formation of UCDs and GCs in general.

\subsection{Individual Cases}
\label{sec:Individual_Cases}

$\textbf{NGC4546 UCD1:}$ The results of our investigation suggest that NGC4546 UCD1 is the 
first unambiguous example of a UCD formed through the stripping of a companion galaxy. Direct 
evidence includes the young measured age of NGC4546 UCD1, an age inconsistent with those 
determined for the vast majority of GCs and UCDs to date. If this UCD had formed through GC 
formation processes, we would expect to see a similarly young GC population, formed alongside 
the UCD. Although our imaging is relatively shallow, we see no clear evidence for a population 
of bright (young) GCs associated with NGC4546. It would also seem extremely unlikely that 
NGC4546 could undergo a recent bout of star and cluster formation, whilst displaying the observed 
universally old ages for the diffuse light of the galaxy. The observed [$\alpha$/Fe] of $\sim$0.0 for 
NGC4546 UCD1 is also interesting, implying a more prolonged period of star formation than typical 
for GCs, which generally have supersolar $\alpha$/Fe \citep[e.g. ][]{Puzia05, Norris08}.

The conclusion that NGC4546 UCD1 is not associated with GC formation processes is additionally 
supported by the observation that the UCD is significantly overluminous relative to the luminosity 
of the host galaxy (see Figure \ref{fig:simulations}). This UCD will remain overluminous even when 
observed at 13~Gyr old, meaning that even if it is a recently formed GGC, it is still extremely difficult 
to explain statistically as a bright outlier of the GCLF. However, it is unlikely to survive as long as 
13~Gyr, as its calculated dynamical friction timescale  is of the order of 0.5~Gyr, consistent with a 
relatively recently formed and short lived object.

Additional evidence for the stripped minor companion origin of the NGC4546 UCD is provided by 
the fact that NGC4546 itself has significant quantities of counterrotating gas \citep{Galletta87,Sage94,Sauron5}. 
This gas rotates in the \textit{same} direction as the UCD but in the opposite direction and slightly 
out of plane relative to the stellar disc of NGC4546. It therefore seems highly likely that NGC4546 
UCD1 is the remaining nucleus of a galaxy stripped within the last few Gyr by NGC4546.

It is possible to estimate the mass of the galaxy that was stripped to form the UCD of NGC4546 
using the observed relation between the nuclear mass of a galaxy (nuclear star cluster + 
supermassive black hole) and the mass of the galaxy spheroid \citep{Graham09}. By combining 
Graham $\&$ Spitler's Eqn. 2 with the measured UCD mass (from our Table \ref{tab:ucd_structural_properties}), 
and assuming that the mass of the supermassive black hole of the progenitor was negligible 
relative to the mass of the nuclear star cluster, we find that the spheroid of the progenitor had 
a stellar mass of $\sim$3.4$^{+1.2}_{-1.5}$$\times$10$^9$~M$_\odot$. This estimate neglects 
the gaseous component of the progenitor, for which we can set a lower limit from the current mass 
of counterrotating gas (assumed to come entirely from the progenitor): M$_{\rm HI}$ = 1.2$\times$10$^8$~M$_\odot$ \citep{Bettoni91} and M$_{\rm H_{2}}$ = 9.6$\times$10$^7$~M$_\odot$  \citep{Sage94}. Based 
on typical gas fractions at the progenitor stellar mass \citep[e.g.][]{Wei10}, we suspect that the 
progenitor originally had much more gas than remains in cold form. 

The estimated progenitor mass is intriguing for two reasons. For one, it is approximately a factor 
of ten less than that of NGC4546 (M $\approx$ 3$\times$10$^{10}$~M$_\odot$, Table \ref{tab:galaxy_properties}), which would allow NGC4546 to disrupt the smaller galaxy without being significantly disordered 
itself~--- reassuring as NGC4546 appears undisturbed. The progenitor mass estimate is also 
interesting because it lies very near the transition between dwarf and giant galaxies (the ``threshold 
mass" discussed in Section \ref{sec:blue_tilts} and \citealt{KGB09}). This result is important because 
as Figure \ref{fig:M87_GCs_UCDs} demonstrates, if it lived long enough, the NGC4546 UCD would 
eventually redden and fade to take up a position in the colour-magnitude diagram consistent with the 
nuclei of giant galaxies. The UCD only lies on the blue colour-magnitude locus currently because it is 
young. Thus the progenitor seems to have had just enough mass to generate a metal-rich nucleus, but 
not enough that a large spheroid or more violent merger would have been expected.
\\
\\
\noindent$\textbf{NGC3923 UCDs:}$ A wealth of circumstantial evidence points to the conclusion 
that the NGC3923 UCDs are GGCs. The existence of multiple UCDs with similar properties (e.g. size) 
would seem unlikely if they were to be the result of several minor merger events. Moreover we appear 
to see the beginnings of transition objects, in the form of five GCs spectroscopically confirmed by 
\citet[][2010, in prep.]{Norris08}, which have luminosities and likely radii that place them between the 
$\sim$3~pc GCs and the $\sim$13~pc UCDs of NGC3923. This behaviour is very similar to that seen 
by \cite{Taylor10} for the more massive GCs of CenA, which similarly appear to bridge the gap between 
GCs and UCDs. Also interesting is the fact that three of the five transitional GCs in NGC3923 have 
measured stellar populations \citep[from][]{Norris08} that place them squarely in the normal GC regime. 
They are all old (age greater than 12~Gyr), with intermediate to high metallicity ([Z/H] from --0.4 to 0.2), 
and supersolar [$\alpha$/Fe]. Although direct stellar population measurements are not yet available 
for the UCDs, evidence of a close connection between the UCDs and GCs of NGC3923 is provided by 
the fact that the UCDs smoothly extend the observed blue tilt in the colour-magnitude relation of the 
blue GCs, suggesting a common mass-metallicity relation.

A GGC origin for the NGC3923 UCDs is also consistent with the computed dynamical friction decay 
timescales. These are all long (t$_{\rm decay}$ $>$ 7~Gyr), implying that the UCDs could be as long 
lived as the GCs of NGC3923, which have ages $>$ 10~Gyr \citep{Norris08}. A final persuasive piece 
of evidence for a connection between the GCs and UCDs of NGC3923 is provided by the fact that the 
UCDs are statistically consistent with being the luminous extension of the observed GC luminosity 
function.
\\
\\
\noindent$\textbf{Sombrero UCD:}$ As discussed in \citet{Hau09} the Sombrero UCD is old (12.6~Gyr), 
metal rich ([Fe/H] = --0.08), and has a slightly supersolar alpha-element enhancement 
([$\alpha$/Fe] = 0.06). These stellar population parameters are consistent with those determined for 
members of the GC population of the Sombrero by \cite{Larsen02b}, as well as being close to that 
measured for the centre of the Sombrero itself by \cite{SanchezBlazquez06b}. Its calculated dynamical 
friction timescale is also long (longer than a Hubble time), indicating it is likely a long lived object. 
When combined with the fact that this UCD can be explained statistically as the massive limit of the 
Sombrero GC system (see Figure \ref{fig:simulations}), it would seem most likely that this UCD has 
a GGC origin. The fact that there are currently no known objects with luminosities between that of the
UCD and the most luminous GCs of the Sombrero may simply reflect the relatively incomplete nature 
of the spectroscopic coverage of the Sombrero GC system (Hau, private communication). However, 
despite the compelling circumstantial evidence it is impossible to be certain that this UCD is a GGC. 
In particular, this UCD could be the result of the stripping of a late-type companion galaxy, if the 
stripping occurred early in the lifetime of the Sombrero on an orbit that allowed the nucleus survive 
unusually long. It could also be the result of the stripping of an early-type dwarf with an older nucleus 
in the more recent past, a plausible scenario given that the Sombrero is known to have some apparently 
tidal features \citep{Malin97}. Possible direct evidence for a stripping origin for the Sombrero UCD is 
provided by the fact that the generalised King, King, and Sersic models for this UCD all show leftover 
low surface brightness features, components that are not seen in the other UCDs (see Figure \ref{fig:ngc3923_ucds}).

\subsection{Integrating UCD and GC Formation}
\label{sec:ucd_formation_in_general}

\begin{figure} 
   \centering
   \begin{turn}{90}
   \includegraphics[scale=0.9]{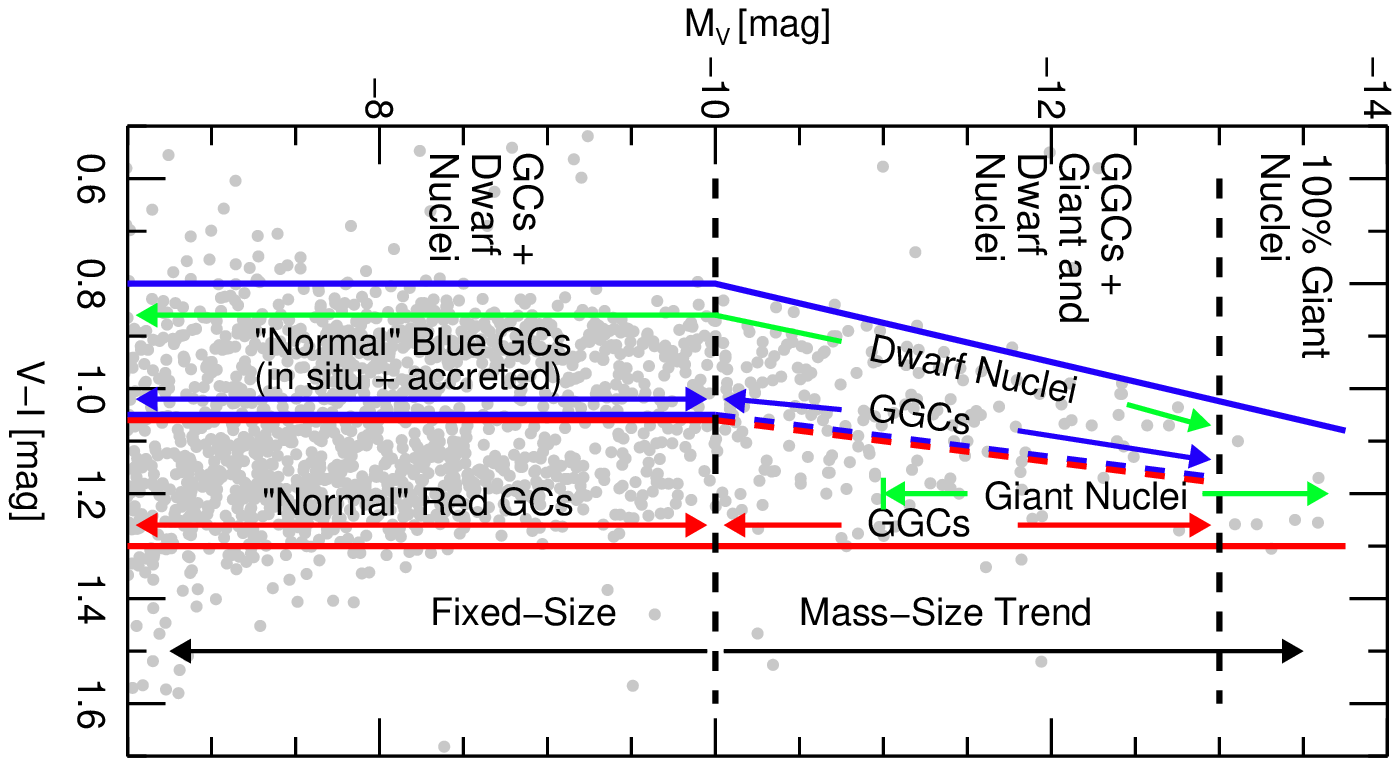}
   \end{turn} 
   \caption{Illustration of a possible scheme describing the origin of GCs and UCDs. The light grey 
   circles are all GCs, UCDs and dwarf nuclei presented in Figure \ref{fig:M87_GCs_UCDs}. ``Normal" 
   GCs refers to those GCs that formed from the collapse of a single giant molecular cloud, and that 
   display a fixed radius with mass trend (i.e. have M $<$ 2$\times$ 10$^6$~M$_\odot$). UCDs are 
   divided into two subgroups, those that form in a GC type manner (i.e. from giant molecular clouds 
   that may or may not merge), which we call GGCs, and those that are the result of stripping the 
   nucleus from a galaxy. Depending on the luminosity of the accreted nucleus, it could be classed as 
   a UCD or mistaken for a normal (usually blue) GC. The black dashed line at M$ _{\rm V}$ = --10 
   indicates the approximate location of the ``scaling onset luminosity", the black dashed line at 
   M$_{\rm V}$ = --13 indicates the luminosity above which UCDs are expected to be exclusively 
   stripped nuclei. The dashed red and blue lines indicate the uncertain division between red and 
   blue GCs/UCDs/nuclei at higher luminosities, due to the increasing effects of self-enrichment on 
   the colours of ``blue" GCs/UCDs/nuclei.}
   \label{fig:scheme}
\end{figure}

\begin{figure} 
   \centering
   \begin{turn}{0}
   \includegraphics[scale=0.9]{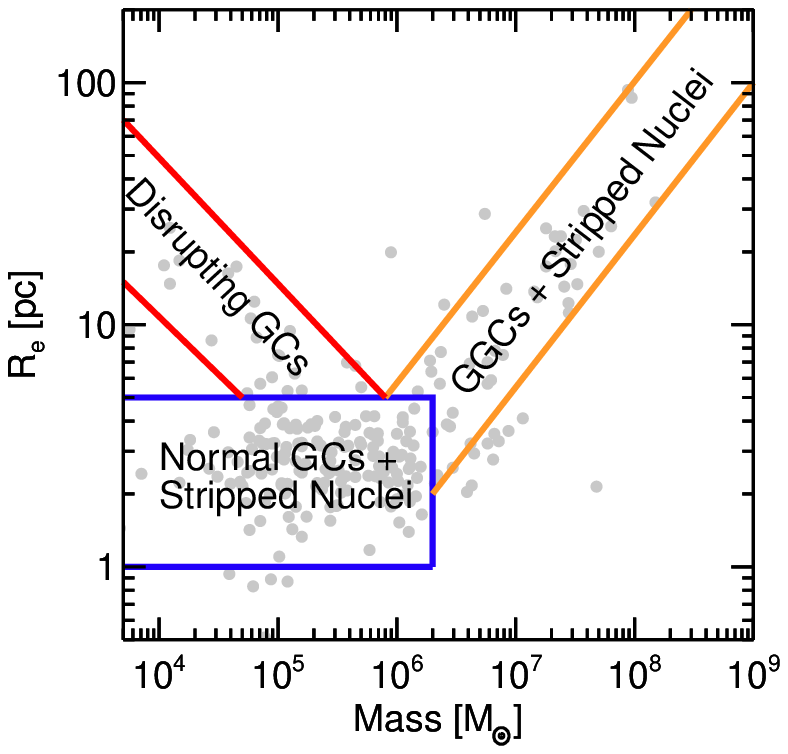}
   \end{turn} 
   \caption{Cartoon schematic showing the origin of various objects in the mass-size plane. UCDs are 
   divided into GGCs, formed either singly or through merging, and stripped galaxy nuclei. Those GCs 
   appearing as ``normal" GCs (either formed in situ or accreted) are contaminated by lower mass 
   stripped galaxy nuclei.}
   \label{fig:ms_scheme}
\end{figure}

Overall, our observations support the idea that UCDs are a ``mixed bag" of objects 
\citep{Hilker09b,Taylor10,DaRocha10}, composed of GGCs and stripped nuclei. We are also confident 
that the GC systems of galaxies are themselves mixed bags, composed of several types of object with 
different formation histories: 1) normal GCs formed in situ, i.e. formed in the galaxies in which they 
currently reside by any of the proposed GC formation scenarios such as dissipational collapse 
\citep{Forbes97} or merger induced star formation \citep{Ashman92}; 2) GCs accreted from smaller 
companions; 3) a population of lower mass stripped dwarf nuclei. In Figures~\ref{fig:scheme} and 
\ref{fig:ms_scheme} we present an illustration of a possible GC, UCD, and nucleus formation scheme 
consistent with this interpretation.

In our proposed scenario, ``normal" GCs form as single collapse clusters, while some, perhaps most, 
UCDs form alongside their host galaxy GC system as GGCs. The GGC-type UCDs are differentiated 
observationally from GCs by the onset of a mass-size trend for cluster mass greater than the scaling 
onset mass of 2$\times$10$^6$~M$_\odot$ \citep[the same break point found in the mass-size relation by e.g.][]{Hasegan05,Mieske08,Taylor10}. The observed change in mass-size behaviour between GC and 
GGC is either the result of a change in the opacity of star forming regions at the scaling onset mass 
\citep[e.g.][and Section \ref{sec:ucd_masses}]{Murray09}, or the result of hierarchical merging of 
multiple YMCs within a single star forming complex. In either of these cases the stellar populations 
of the resulting GGC-type UCDs would largely indistinguishable from those of other lower mass normal 
GCs that escaped from the same star forming complexes without merging. In particular, even for UCDs 
formed by merging, the timescale of  star formation and merging for each UCD should be relatively 
rapid ($<$  a few hundred~Myr). Such a short star forming epoch leads to supersolar alpha-element 
enhancement, as is usually found in normal GCs \citep[e.g.][]{Puzia05,Norris08}.

How far this hierarchical process may proceed is uncertain; building the most massive UCDs would 
likely require the merger of hundreds of normal GCs, a process that seems somewhat implausible. 
This would therefore seem to imply the existence of an upper limit to the size of the UCDs that this 
process could create, at least in practice. From purely statistical arguments (see Section \ref{sec:gclf_analysis}), 
and assuming the number of merged systems scales as the number of unmerged GCs, we suggest 
that 7$\times$10$^7$~M$_\odot$ (M$_{\rm V}$ = --13 for old systems) represents the practical upper 
limit to UCD formation by GC-like processes. GC systems sufficiently rich to populate above this limit 
(either by merging GCs or by the collapse of single giant clouds with changed physics) are not known 
at the present time. We note that our suggested upper limit of 7$\times$10$^7$~M$_\odot$ is also very 
similar to the mass of stars formed in superclusters (clusters of star clusters) seen in some interacting 
and strongly star forming galaxies. \citet{Fellhauer05} model the evolution of such structures and find 
that they can indeed merge over short ($<$  a few hundred~Myr) timescales to form UCD-like objects 
such as the young ($\sim$500~Myr), massive (dynamical mass $\sim$ 8$\times$10$^7$~M$_\odot$) 
cluster W3 associated with the merger remnant NGC7252 \citep{Maraston04}.

On top of this population of GCs and GGC-type UCDs intrinsic to the host galaxy, there are two 
sets of interloper stellar systems. The first is simply the population of normal GCs provided to larger 
galaxies by accreted galaxies \citep[these are mostly blue GCs, as low mass galaxies have very few 
red GCs, e.g.,][]{Lotz04,Peng06}. This population can be a significant fraction of a galaxy's GC system. 
Recent work on the Milky Way GC system shows that perhaps 25$\%$ of its GCs could have been 
provided by 6-8 dwarf galaxies \citep{Forbes10}. Similarly \cite{Mackey10} find that $\ga$80$\%$ of 
Andromeda's outer GCs are likely to have been accreted from dwarf satellites.

The second (presumably smaller) population of interlopers is provided by the nuclei of the accreted 
galaxies. If a stripped nucleus has M $<$ 2$\times$10$^6$~M$_\odot$ (and has a small radius), it is 
possible for it to be lost amongst the larger population of normal GCs, at least in colour-magnitude and 
mass-size space. In principle an in-depth examination of the object's stellar populations may provide 
clues to its true origin, especially if its [$\alpha$/Fe] ratio is depressed relative to GCs, as galaxy nuclei 
are expected to have longer formation timescales than GCs/GGCs. It has been suggested that such 
remnant nuclei include $\omega$~Centauri in the Milky Way \citep[e.g.][]{Hilker00,Mackey05} and G1 
around the Andromeda galaxy \citep[e.g.][]{Meylan01}. Alternatively, if the remnant nucleus has M $>$
2$\times$10$^6$~M$_\odot$ it will be classified as a UCD. The UCD of NGC4546 is the first conclusive 
example of such a stripped-nucleus UCD.

Such stripped-nucleus UCDs can also explain the existence of UCDs above the apparent upper 
magnitude limit for GGC-type UCDs at M$_{\rm V}$ $\sim$ --13. Interestingly, examination of 
Figure \ref{fig:M87_GCs_UCDs} demonstrates that for these most extreme objects, it is not in fact 
dwarf galaxies (M$_{\rm B}$ $>$ --17.6, V$_{\rm circ}$ $<$ 120~kms$^{-1}$) that must be stripped, 
but likely giant galaxies (though perhaps ``small giants" as in the NGC4546 case; see Section \ref{sec:Individual_Cases}).
As well as being consistent with the most luminous UCDs, the galaxy nuclei observed in group/cluster 
samples have the correct colour-magnitude trends to explain any blue GC/UCD as well as the more 
luminous red UCDs. At first glance it would seem that red objects with M$_{\rm V}$ fainter than $\sim$--11 
(the magnitude of the lowest luminosity red nuclei in Figure \ref{fig:M87_GCs_UCDs}) must form only 
through GC-like processes. However, it is possible that red nuclei (from giant galaxies), or massive 
blue nuclei with colours as red as low-mass red GCs (from relatively high mass dwarf galaxies), may 
lose much of their mass during the stripping event and during their subsequent dynamical evolution 
in the larger galaxy potential. This stripping/disruption could allow objects that started out with M$_{\rm V}$ 
more luminous than --11 to eventually fade into the lower luminosity region mostly inhabited by red 
GGCs.

\section{Conclusions}

Using SOAR/Goodman imaging and spectroscopy, SAURON IFU spectroscopy, and archival \textit{HST} 
ACS and WFPC2 imaging, we have investigated a sample of field/group UCDs. Our main conclusions 
are as follows:

\begin{itemize}

\item To first order we find that low density environment UCDs are as common as UCDs in galaxy 
	clusters.

\item Combining our UCD sample with literature data we confirm that UCDs are a ``mixed bag" 
	\citep{Hilker09b,Taylor10,DaRocha10}, composed of objects we term giant-GCs (GGCs), 
	and stripped galaxy nuclei. 
	
\item In the NGC4546 UCD we find the first unambiguous example of a young UCD formed by 
	the stripping of a galaxy nucleus. The UCD of NGC4546 is young ($\sim$3.4~Gyr), metal 
	rich, and has near solar alpha-element abundance ratios. In stark contrast the diffuse light 
	of NGC4546 is old ($\sim$10 Gyr), intermediate in metallicity, and has supersolar [$\alpha$/Fe]. 
	Additionally the NGC4546 UCD is significantly overluminous relative to the host galaxy, 
	and hence cannot be explained as the bright extension of the GCLF of NGC4546. The short 
	dynamical friction decay timescale of 0.5~Gyr of this UCD also indicates that it is likely a 
	short lived object.
	
\item We have discovered multiple UCDs (at least two, possibly as many as eight) associated 
	with the group shell elliptical NGC3923. This multiplicity of UCDs, along with the fact that 
	the UCDs all have long dynamical friction decay timescales, and that they are consistent 
	with the extrapolation of the GCLF and colour-magnitude diagram of NGC3923, indicates 
	that these UCDs are GGCs.	
	
\item The UCD of the Sombrero galaxy is likely a GGC but a stripping origin cannot be ruled out. 
	Its stellar populations are consistent with those of the GC system of the Sombrero, and its 
	luminosity is explainable as the bright end of the GCLF of the Sombrero. Additionally it has 
	a long dynamical friction decay timescale ($>$ 25 Gyr).	
	
\item We find that the colour-magnitude diagrams of dwarf nuclei, blue GCs, and blue UCDs are 
	indistinguishable. Dwarf nuclei exist over the entire range of blue GC and UCD luminosities. 
	All three populations display the same ``blue tilt" phenomenon (a mass-metallicity relation). 
	Red GCs, red UCDs, and more massive dwarf/giant galaxy nuclei also show considerable 
	overlap in colour-magnitude space but display no evidence for a mass-metallicity relation.
	
\item We suggest the existence of a common ``scaling onset mass" of $\sim$2$\times$10$^6$~M$_\odot$
	marking the onset of both a mass-size relation (defining UCDs as a class) and a 
	mass-metallicity relation (the blue tilt). In line with previous studies we find that UCDs (of 
	either GGC or stripped nucleus type) may be efficiently selected by requiring objects to 
	have R$_{\rm e}$ $>$ 6~pc and M$_{\rm V}$ $<$ --10 or mass $>$ 2$\times$10$^6$~M$_\odot$.
	
\item From statistical arguments we suggest M$_{\rm V}$ $<$ --13 as the upper luminosity limit 
	for an old UCD formed via the GGC route. GC systems sufficiently populous to create GGCs 
	above this limit do not exist at the present epoch.	

\item We present a general scheme that unifies the formation of GCs, UCDs (both GGCs and 
	stripped nuclei), and galaxy nuclei. We argue that the evidence points to a scenario in 
	which the UCDs of NGC3923 and perhaps the majority of UCDs in general are GGCs. 
	These GGCs form alongside normal GCs (perhaps even from the merging of normal GCs) 
	during the major epochs of spheroid assembly. Superimposed on this population is a second 
	population of UCDs produced by the stripping of nuclei from galaxies. These objects are 
	most easily identified in cases where the resulting UCD is too luminous to be explained 
	as the bright end of the GCLF (as in the case of the NGC4546 UCD, Fornax UCD3, and 
	Virgo VUCD7). Such objects could also occur at fainter magnitudes, hidden by the 
	indistinguishable colours of blue GCs/UCDs and dwarf nuclei. 

\end{itemize}

The existence of two populations of Ultra Compact Dwarfs, one related to the major epochs of 
spheroid assembly and one to minor merger events, is clearly of great utility in the study of galaxy 
formation. In order for UCDs to fullfill their potential as probes of the galaxy formation process, 
it will be necessary to be able to more cleanly separate the two distinct UCD types from each 
other. To help achieve this goal we are now undertaking a comprehensive survey of all available 
\textit{HST} imaging and following up with spectroscopy to find UCDs located in all environments. 
This large sample will allow us to examine the demographics of each UCD type and probe the 
implications for minor and major merger rates as a function of environment.

\section{Acknowledgements}

We thank the anonymous referee for comments that greatly improved this manuscript. We thank 
Brad Barlow, Bart Dunlap and Chris Clemens for assistance in the design, manufacture and use 
of the NGC3923 SOAR/Goodman MOS masks. We thank Harald Kuntschner for his help with the 
analysis of the SAURON observations of the NGC4546 UCD, as well as Bob O'Connell and 
George Hau for interesting discussions that have improved this manuscript.

We acknowledge the Goodman family for providing the financial support necessary for the 
construction of the Goodman Spectrograph. We are also grateful for the support provided by the 
SOAR operators: Alberto Pasten, Patricio Ugarte, Sergio Pizarro, and Daniel Maturana. The SOAR 
Telescope is operated by the Association of Universities for Research in Astronomy, Inc., under a 
cooperative agreement between the CNPq, Brazil, the National Observatory for Optical Astronomy 
(NOAO), the University of North Carolina, and Michigan State University, USA. 

Based on observations made with the NASA/ESA Hubble Space Telescope, obtained from the 
data archive at the Space Telescope Science Institute. STScI is operated by the Association of 
Universities for Research in Astronomy, Inc. under NASA contract NAS 5-26555.

We thank the SAURON collaboration for providing their datacube for NGC4546. The SAURON 
project is made possible through grants 614.13.003 and 781.74.203 from ASTRON/NWO and 
financial contributions from the Institut National des Sciences de l'Univers, the Universite Claude 
Bernard Lyon I, the universities of Durham and Leiden, the British Council, PPARC grant 
`Extragalactic Astronomy $\&$ Cosmology at Durham 1998-2002', and the Netherlands Research 
School for Astronomy NOVA. 

This publication makes use of data products from the Two Micron All Sky Survey, which is a joint 
project of the University of Massachusetts and the Infrared Processing and Analysis Center/California 
Institute of Technology, funded by the National Aeronautics and Space Administration and the 
National Science Foundation.

\bibliographystyle{mn2e}
\bibliography{references}

\begin{thebibliography}{126}
\expandafter\ifx\csname natexlab\endcsname\relax\def\natexlab#1{#1}\fi

\bibitem[{{Agarwal} \& {Milosavljevic}(2010)}]{Agarwal10}
{Agarwal} M., {Milosavljevic} M., 2010, ArXiv e-prints

\bibitem[{{Ashman} {et~al.}(1994{\natexlab{a}}){Ashman}, {Bird}, \&
  {Zepf}}]{AshmanBirdZepf}
{Ashman} K.~M., {Bird} C.~M., {Zepf} S.~E., 1994{\natexlab{a}}, \aj, 108, 2348

\bibitem[{{Ashman} {et~al.}(1994{\natexlab{b}}){Ashman}, {Bird}, \&
  {Zepf}}]{KMM}
---, 1994{\natexlab{b}}, \aj, 108, 2348

\bibitem[{{Ashman} \& {Zepf}(1992)}]{Ashman92}
{Ashman} K.~M., {Zepf} S.~E., 1992, \apj, 384, 50

\bibitem[{{Bacon} {et~al.}(2001){Bacon}, {Copin}, {Monnet}, {Miller},
  {Allington-Smith}, {Bureau}, {Carollo}, {Davies}, {Emsellem}, {Kuntschner},
  {Peletier}, {Verolme}, \& {de Zeeuw}}]{SAURON1}
{Bacon} R., {Copin} Y., {Monnet} G., {Miller} B.~W., {Allington-Smith} J.~R.,
  {Bureau} M., {Carollo} C.~M., {Davies} R.~L., {Emsellem} E., {Kuntschner} H.,
  {Peletier} R.~F., {Verolme} E.~K., {de Zeeuw} P.~T., 2001, \mnras, 326, 23

\bibitem[{{Bailin} \& {Harris}(2009)}]{Bailin_Harris09}
{Bailin} J., {Harris} W.~E., 2009, \apj, 695, 1082

\bibitem[{{Barmby} {et~al.}(2007){Barmby}, {McLaughlin}, {Harris}, {Harris}, \&
  {Forbes}}]{Barmby07}
{Barmby} P., {McLaughlin} D.~E., {Harris} W.~E., {Harris} G.~L.~H., {Forbes}
  D.~A., 2007, \aj, 133, 2764

\bibitem[{{Bekki} {et~al.}(2001){Bekki}, {Couch}, \& {Drinkwater}}]{Bekki01}
{Bekki} K., {Couch} W.~J., {Drinkwater} M.~J., 2001, \apjl, 552, L105

\bibitem[{{Bekki} {et~al.}(2003){Bekki}, {Couch}, {Drinkwater}, \&
  {Shioya}}]{Bekki03c}
{Bekki} K., {Couch} W.~J., {Drinkwater} M.~J., {Shioya} Y., 2003, \mnras, 344,
  399

\bibitem[{{Bekki} \& {Freeman}(2003)}]{Bekki03b}
{Bekki} K., {Freeman} K.~C., 2003, \mnras, 346, L11

\bibitem[{{Bell} \& {de Jong}(2001)}]{Bell01}
{Bell} E.~F., {de Jong} R.~S., 2001, \apj, 550, 212

\bibitem[{{Bertin} \& {Arnouts}(1996)}]{Sextractor96}
{Bertin} E., {Arnouts} S., 1996, \aaps, 117, 393

\bibitem[{{Bettoni} {et~al.}(1991){Bettoni}, {Galletta}, \&
  {Oosterloo}}]{Bettoni91}
{Bettoni} D., {Galletta} G., {Oosterloo} T., 1991, \mnras, 248, 544

\bibitem[{{Binney} \& {Tremaine}(2008)}]{BinneyTremaine08}
{Binney} J., {Tremaine} S., 2008, {Galactic Dynamics: Second Edition}, {Binney,
  J.~\& Tremaine, S.}, ed. Princeton University Press

\bibitem[{{Brown} {et~al.}(2003){Brown}, {Forbes}, {Silva}, {Helsdon},
  {Ponman}, {Hau}, {Brodie}, {Goudfrooij}, \& {Bothun}}]{Brown03}
{Brown} R.~J.~N., {Forbes} D.~A., {Silva} D., {Helsdon} S.~F., {Ponman} T.~J.,
  {Hau} G.~K.~T., {Brodie} J.~P., {Goudfrooij} P., {Bothun} G., 2003, \mnras,
  341, 747

\bibitem[{{Bruzual} \& {Charlot}(2003)}]{BruzualCharlot}
{Bruzual} G., {Charlot} S., 2003, \mnras, 344, 1000

\bibitem[{{Cappellari} \& {Emsellem}(2004)}]{ppxf}
{Cappellari} M., {Emsellem} E., 2004, \pasp, 116, 138

\bibitem[{{Chilingarian} {et~al.}(2010){Chilingarian}, {Mieske}, {Hilker}, \&
  {Infante}}]{Chilingarian10}
{Chilingarian} I., {Mieske} S., {Hilker} M., {Infante} L., 2010, ArXiv e-prints

\bibitem[{{Chilingarian} {et~al.}(2008){Chilingarian}, {Cayatte}, \&
  {Bergond}}]{Chilingarian08b}
{Chilingarian} I.~V., {Cayatte} V., {Bergond} G., 2008, \mnras, 390, 906

\bibitem[{{Chilingarian} \& {Mamon}(2008)}]{Chilingarian08a}
{Chilingarian} I.~V., {Mamon} G.~A., 2008, \mnras, 385, L83

\bibitem[{{Cho}(2008)}]{ChoThesis}
{Cho} J., 2008, PhD thesis, University of Durham

\bibitem[{{Clemens} {et~al.}(2004){Clemens}, {Crain}, \&
  {Anderson}}]{GoodmanSpec}
{Clemens} J.~C., {Crain} J.~A., {Anderson} R., 2004, in Society of
  Photo-Optical Instrumentation Engineers (SPIE) Conference Series, Vol. 5492,
  Society of Photo-Optical Instrumentation Engineers (SPIE) Conference Series,
  {A.~F.~M.~Moorwood \& M.~Iye}, ed., pp. 331--340

\bibitem[{{C{\^o}t{\'e}} {et~al.}(2006){C{\^o}t{\'e}}, {Piatek}, {Ferrarese},
  {Jord{\'a}n}, {Merritt}, {Peng}, {Ha{\c s}egan}, {Blakeslee}, {Mei}, {West},
  {Milosavljevi{\'c}}, \& {Tonry}}]{Cote06}
{C{\^o}t{\'e}} P., {Piatek} S., {Ferrarese} L., {Jord{\'a}n} A., {Merritt} D.,
  {Peng} E.~W., {Ha{\c s}egan} M., {Blakeslee} J.~P., {Mei} S., {West} M.~J.,
  {Milosavljevi{\'c}} M., {Tonry} J.~L., 2006, \apjs, 165, 57

\bibitem[{{Da Rocha} {et~al.}(2010){Da Rocha}, {Mieske}, {Georgiev}, {Hilker},
  {Ziegler}, \& {Mendes de Oliveira}}]{DaRocha10}
{Da Rocha} C., {Mieske} S., {Georgiev} I.~Y., {Hilker} M., {Ziegler} B.~L.,
  {Mendes de Oliveira} C., 2010, ArXiv e-prints

\bibitem[{{Dabringhausen} {et~al.}(2008){Dabringhausen}, {Hilker}, \&
  {Kroupa}}]{Dabringhausen08}
{Dabringhausen} J., {Hilker} M., {Kroupa} P., 2008, \mnras, 386, 864

\bibitem[{{Dalcanton} {et~al.}(2004){Dalcanton}, {Yoachim}, \&
  {Bernstein}}]{Dalcanton04}
{Dalcanton} J.~J., {Yoachim} P., {Bernstein} R.~A., 2004, \apj, 608, 189

\bibitem[{{Drinkwater} {et~al.}(2000){Drinkwater}, {Jones}, {Gregg}, \&
  {Phillipps}}]{Drinkwater00}
{Drinkwater} M.~J., {Jones} J.~B., {Gregg} M.~D., {Phillipps} S., 2000,
  Publications of the Astronomical Society of Australia, 17, 227

\bibitem[{{Elson}(1999)}]{Elson99}
{Elson} R.~A.~W., 1999, in Globular Clusters, {C.~Mart{\'{\i}}nez Roger,
  I.~Perez Fourn{\'o}n, \& F.~S{\'a}nchez }, ed., pp. 209--248

\bibitem[{{Evstigneeva} {et~al.}(2007{\natexlab{a}}){Evstigneeva},
  {Drinkwater}, {Jurek}, {Firth}, {Jones}, {Gregg}, \&
  {Phillipps}}]{Evstigneeva07b}
{Evstigneeva} E.~A., {Drinkwater} M.~J., {Jurek} R., {Firth} P., {Jones} J.~B.,
  {Gregg} M.~D., {Phillipps} S., 2007{\natexlab{a}}, \mnras, 378, 1036

\bibitem[{{Evstigneeva} {et~al.}(2008){Evstigneeva}, {Drinkwater}, {Peng},
  {Hilker}, {De Propris}, {Jones}, {Phillipps}, {Gregg}, \&
  {Karick}}]{Evstigneeva08}
{Evstigneeva} E.~A., {Drinkwater} M.~J., {Peng} C.~Y., {Hilker} M., {De
  Propris} R., {Jones} J.~B., {Phillipps} S., {Gregg} M.~D., {Karick} A.~M.,
  2008, \aj, 136, 461

\bibitem[{{Evstigneeva} {et~al.}(2007{\natexlab{b}}){Evstigneeva}, {Gregg},
  {Drinkwater}, \& {Hilker}}]{Evstigneeva07a}
{Evstigneeva} E.~A., {Gregg} M.~D., {Drinkwater} M.~J., {Hilker} M.,
  2007{\natexlab{b}}, \aj, 133, 1722

\bibitem[{{Fellhauer} \& {Kroupa}(2002)}]{Fellhauer02}
{Fellhauer} M., {Kroupa} P., 2002, \mnras, 330, 642

\bibitem[{{Fellhauer} \& {Kroupa}(2005)}]{Fellhauer05}
---, 2005, \mnras, 359, 223

\bibitem[{{Firth} {et~al.}(2007){Firth}, {Drinkwater}, {Evstigneeva}, {Gregg},
  {Karick}, {Jones}, \& {Phillipps}}]{Firth07}
{Firth} P., {Drinkwater} M.~J., {Evstigneeva} E.~A., {Gregg} M.~D., {Karick}
  A.~M., {Jones} J.~B., {Phillipps} S., 2007, \mnras, 382, 1342

\bibitem[{{Firth} {et~al.}(2008){Firth}, {Drinkwater}, \& {Karick}}]{Firth08}
{Firth} P., {Drinkwater} M.~J., {Karick} A.~M., 2008, \mnras, 389, 1539

\bibitem[{{Forbes} \& {Bridges}(2010)}]{Forbes10}
{Forbes} D.~A., {Bridges} T., 2010, \mnras, 404, 1203

\bibitem[{{Forbes} {et~al.}(1997){Forbes}, {Brodie}, \& {Grillmair}}]{Forbes97}
{Forbes} D.~A., {Brodie} J.~P., {Grillmair} C.~J., 1997, \aj, 113, 1652

\bibitem[{{Forbes} {et~al.}(2010){Forbes}, {Spitler}, {Harris}, {Bailin},
  {Strader}, {Brodie}, \& {Larsen}}]{Forbes10b}
{Forbes} D.~A., {Spitler} L.~R., {Harris} W.~E., {Bailin} J., {Strader} J.,
  {Brodie} J.~P., {Larsen} S.~S., 2010, \mnras, 403, 429

\bibitem[{{Galletta}(1987)}]{Galletta87}
{Galletta} G., 1987, \apj, 318, 531

\bibitem[{{Garnett}(2002)}]{Garnett02}
{Garnett} D.~R., 2002, \apj, 581, 1019

\bibitem[{{Gieles} {et~al.}(2010){Gieles}, {Baumgardt}, {Heggie}, \&
  {Lamers}}]{Gieles10}
{Gieles} M., {Baumgardt} H., {Heggie} D., {Lamers} H., 2010, ArXiv e-prints

\bibitem[{{Giuricin} {et~al.}(2000){Giuricin}, {Marinoni}, {Ceriani}, \&
  {Pisani}}]{Giuricin00}
{Giuricin} G., {Marinoni} C., {Ceriani} L., {Pisani} A., 2000, \apj, 543, 178

\bibitem[{{Goerdt} {et~al.}(2008){Goerdt}, {Moore}, {Kazantzidis}, {Kaufmann},
  {Macci{\`o}}, \& {Stadel}}]{Goerdt08}
{Goerdt} T., {Moore} B., {Kazantzidis} S., {Kaufmann} T., {Macci{\`o}} A.~V.,
  {Stadel} J., 2008, \mnras, 385, 2136

\bibitem[{{Graham} \& {Spitler}(2009)}]{Graham09}
{Graham} A.~W., {Spitler} L.~R., 2009, \mnras, 397, 2148

\bibitem[{{Gregg} {et~al.}(2009){Gregg}, {Drinkwater}, {Evstigneeva}, {Jurek},
  {Karick}, {Phillipps}, {Bridges}, {Jones}, {Bekki}, \& {Couch}}]{Gregg09}
{Gregg} M.~D., {Drinkwater} M.~J., {Evstigneeva} E., {Jurek} R., {Karick}
  A.~M., {Phillipps} S., {Bridges} T., {Jones} J.~B., {Bekki} K., {Couch}
  W.~J., 2009, \aj, 137, 498

\bibitem[{{Ha{\c s}egan} {et~al.}(2005){Ha{\c s}egan}, {Jord{\'a}n},
  {C{\^o}t{\'e}}, {Djorgovski}, {McLaughlin}, {Blakeslee}, {Mei}, {West},
  {Peng}, {Ferrarese}, {Milosavljevi{\'c}}, {Tonry}, \& {Merritt}}]{Hasegan05}
{Ha{\c s}egan} M., {Jord{\'a}n} A., {C{\^o}t{\'e}} P., {Djorgovski} S.~G.,
  {McLaughlin} D.~E., {Blakeslee} J.~P., {Mei} S., {West} M.~J., {Peng} E.~W.,
  {Ferrarese} L., {Milosavljevi{\'c}} M., {Tonry} J.~L., {Merritt} D., 2005,
  \apj, 627, 203

\bibitem[{{Harris}(2009)}]{Harris09}
{Harris} W.~E., 2009, \apj, 699, 254

\bibitem[{{Harris} {et~al.}(2002){Harris}, {Harris}, {Holland}, \&
  {McLaughlin}}]{Harris02}
{Harris} W.~E., {Harris} G.~L.~H., {Holland} S.~T., {McLaughlin} D.~E., 2002,
  \aj, 124, 1435

\bibitem[{{Harris} {et~al.}(2006){Harris}, {Whitmore}, {Karakla}, {Oko{\'n}},
  {Baum}, {Hanes}, \& {Kavelaars}}]{Harris06}
{Harris} W.~E., {Whitmore} B.~C., {Karakla} D., {Oko{\'n}} W., {Baum} W.~A.,
  {Hanes} D.~A., {Kavelaars} J.~J., 2006, \apj, 636, 90

\bibitem[{{Hau} {et~al.}(2009){Hau}, {Spitler}, {Forbes}, {Proctor}, {Strader},
  {Mendel}, {Brodie}, \& {Harris}}]{Hau09}
{Hau} G.~K.~T., {Spitler} L.~R., {Forbes} D.~A., {Proctor} R.~N., {Strader} J.,
  {Mendel} J.~T., {Brodie} J.~P., {Harris} W.~E., 2009, \mnras, 394, L97

\bibitem[{{Hilker}(2009{\natexlab{a}})}]{Hilker09}
{Hilker} M., 2009{\natexlab{a}}, ArXiv e-prints:0906.0776

\bibitem[{{Hilker}(2009{\natexlab{b}})}]{Hilker09b}
---, 2009{\natexlab{b}}, {UCDs - A Mixed Bag of Objects}, {Richtler, T.~\&
  Larsen, S.}, ed., pp. 51--+

\bibitem[{{Hilker} {et~al.}(1999){Hilker}, {Infante}, {Vieira},
  {Kissler-Patig}, \& {Richtler}}]{Hilker99}
{Hilker} M., {Infante} L., {Vieira} G., {Kissler-Patig} M., {Richtler} T.,
  1999, \aaps, 134, 75

\bibitem[{{Hilker} \& {Richtler}(2000)}]{Hilker00}
{Hilker} M., {Richtler} T., 2000, \aap, 362, 895

\bibitem[{{Jensen} {et~al.}(2003){Jensen}, {Tonry}, {Barris}, {Thompson},
  {Liu}, {Rieke}, {Ajhar}, \& {Blakeslee}}]{Jensen03}
{Jensen} J.~B., {Tonry} J.~L., {Barris} B.~J., {Thompson} R.~I., {Liu} M.~C.,
  {Rieke} M.~J., {Ajhar} E.~A., {Blakeslee} J.~P., 2003, \apj, 583, 712

\bibitem[{{Jones} {et~al.}(2006){Jones}, {Drinkwater}, {Jurek}, {Phillipps},
  {Gregg}, {Bekki}, {Couch}, {Karick}, {Parker}, \& {Smith}}]{Jones06}
{Jones} J.~B., {Drinkwater} M.~J., {Jurek} R., {Phillipps} S., {Gregg} M.~D.,
  {Bekki} K., {Couch} W.~J., {Karick} A., {Parker} Q.~A., {Smith} R.~M., 2006,
  \aj, 131, 312

\bibitem[{{Jord{\'a}n} {et~al.}(2005){Jord{\'a}n}, {C{\^o}t{\'e}}, {Blakeslee},
  {Ferrarese}, {McLaughlin}, {Mei}, {Peng}, {Tonry}, {Merritt},
  {Milosavljevi{\'c}}, {Sarazin}, {Sivakoff}, \& {West}}]{Jordan05}
{Jord{\'a}n} A., {C{\^o}t{\'e}} P., {Blakeslee} J.~P., {Ferrarese} L.,
  {McLaughlin} D.~E., {Mei} S., {Peng} E.~W., {Tonry} J.~L., {Merritt} D.,
  {Milosavljevi{\'c}} M., {Sarazin} C.~L., {Sivakoff} G.~R., {West} M.~J.,
  2005, \apj, 634, 1002

\bibitem[{{Jord{\'a}n} {et~al.}(2007){Jord{\'a}n}, {McLaughlin},
  {C{\^o}t{\'e}}, {Ferrarese}, {Peng}, {Mei}, {Villegas}, {Merritt}, {Tonry},
  \& {West}}]{Jordan07}
{Jord{\'a}n} A., {McLaughlin} D.~E., {C{\^o}t{\'e}} P., {Ferrarese} L., {Peng}
  E.~W., {Mei} S., {Villegas} D., {Merritt} D., {Tonry} J.~L., {West} M.~J.,
  2007, \apjs, 171, 101

\bibitem[{{Kannappan}(2004)}]{Kannappan04}
{Kannappan} S.~J., 2004, \apjl, 611, L89

\bibitem[{{Kannappan} {et~al.}(2002){Kannappan}, {Fabricant}, \&
  {Franx}}]{Kannappan02}
{Kannappan} S.~J., {Fabricant} D.~G., {Franx} M., 2002, \aj, 123, 2358

\bibitem[{{Kannappan} \& {Gawiser}(2007)}]{Kannappan07}
{Kannappan} S.~J., {Gawiser} E., 2007, \apjl, 657, L5

\bibitem[{{Kannappan} {et~al.}(2009){Kannappan}, {Guie}, \& {Baker}}]{KGB09}
{Kannappan} S.~J., {Guie} J.~M., {Baker} A.~J., 2009, \aj, 138, 579

\bibitem[{{Kannappan} \& {Wei}(2008)}]{KW08}
{Kannappan} S.~J., {Wei} L.~H., 2008, in American Institute of Physics
  Conference Series, Vol. 1035, The Evolution of Galaxies Through the Neutral
  Hydrogen Window, {R.~Minchin \& E.~Momjian}, ed., pp. 163--168

\bibitem[{{King}(1962)}]{King62}
{King} I., 1962, \aj, 67, 471

\bibitem[{{Krist}(1995)}]{tinytim}
{Krist} J., 1995, in Astronomical Society of the Pacific Conference Series,
  Vol.~77, Astronomical Data Analysis Software and Systems IV, {R.~A.~Shaw,
  H.~E.~Payne, \& J.~J.~E.~Hayes}, ed., pp. 349--+

\bibitem[{{Kundu} \& {Whitmore}(2001{\natexlab{a}})}]{Kundu01a}
{Kundu} A., {Whitmore} B.~C., 2001{\natexlab{a}}, \aj, 121, 2950

\bibitem[{{Kundu} \& {Whitmore}(2001{\natexlab{b}})}]{Kundu01b}
---, 2001{\natexlab{b}}, \aj, 122, 1251

\bibitem[{{Kuntschner}(2004)}]{Kuntschner04}
{Kuntschner} H., 2004, \aap, 426, 737

\bibitem[{{Kuntschner} {et~al.}(2006){Kuntschner}, {Emsellem}, {Bacon},
  {Bureau}, {Cappellari}, {Davies}, {de Zeeuw}, {Falc{\'o}n-Barroso},
  {Krajnovi{\'c}}, {McDermid}, {Peletier}, \& {Sarzi}}]{Kuntschner06}
{Kuntschner} H., {Emsellem} E., {Bacon} R., {Bureau} M., {Cappellari} M.,
  {Davies} R.~L., {de Zeeuw} P.~T., {Falc{\'o}n-Barroso} J., {Krajnovi{\'c}}
  D., {McDermid} R.~M., {Peletier} R.~F., {Sarzi} M., 2006, \mnras, 369, 497

\bibitem[{{Kuntschner} {et~al.}(2010){Kuntschner}, {Emsellem}, {Bacon},
  {Cappellari}, {Davies}, {de Zeeuw}, {Falc{\'o}n-Barroso}, {Krajnovi{\'c}},
  {McDermid}, {Peletier}, {Sarzi}, {Shapiro}, {van den Bosch}, \& {van de
  Ven}}]{Kuntschner10}
{Kuntschner} H., {Emsellem} E., {Bacon} R., {Cappellari} M., {Davies} R.~L.,
  {de Zeeuw} P.~T., {Falc{\'o}n-Barroso} J., {Krajnovi{\'c}} D., {McDermid}
  R.~M., {Peletier} R.~F., {Sarzi} M., {Shapiro} K.~L., {van den Bosch}
  R.~C.~E., {van de Ven} G., 2010, \mnras, 408, 97

\bibitem[{{Landolt}(2009)}]{Landolt09}
{Landolt} A.~U., 2009, \aj, 137, 4186

\bibitem[{{Larsen} \& {Brodie}(2002)}]{Larsen02}
{Larsen} S.~S., {Brodie} J.~P., 2002, \aj, 123, 1488

\bibitem[{{Larsen} {et~al.}(2002){Larsen}, {Brodie}, {Beasley}, \&
  {Forbes}}]{Larsen02b}
{Larsen} S.~S., {Brodie} J.~P., {Beasley} M.~A., {Forbes} D.~A., 2002, \aj,
  124, 828

\bibitem[{{Lotz} {et~al.}(2004){Lotz}, {Miller}, \& {Ferguson}}]{Lotz04}
{Lotz} J.~M., {Miller} B.~W., {Ferguson} H.~C., 2004, \apj, 613, 262

\bibitem[{{Mackey} \& {van den Bergh}(2005)}]{Mackey05}
{Mackey} A.~D., {van den Bergh} S., 2005, \mnras, 360, 631

\bibitem[{{Mackey} {et~al.}(2010){Mackey}, {Huxor}, {Ferguson}, {Irwin},
  {Tanvir}, {McConnachie}, {Ibata}, {Chapman}, \& {Lewis}}]{Mackey10}
{Mackey} D., {Huxor} A., {Ferguson} A., {Irwin} M., {Tanvir} N., {McConnachie}
  A., {Ibata} R., {Chapman} S., {Lewis} G., 2010, ArXiv e-prints

\bibitem[{{Madrid} {et~al.}(2010){Madrid}, {Graham}, {Harris}, {Goudfrooij},
  {Forbes}, {Carter}, {Blakeslee}, {Spitler}, \& {Ferguson}}]{Madrid10}
{Madrid} J.~P., {Graham} A.~W., {Harris} W.~E., {Goudfrooij} P., {Forbes}
  D.~A., {Carter} D., {Blakeslee} J.~P., {Spitler} L.~R., {Ferguson} H.~C.,
  2010, ArXiv e-prints

\bibitem[{{Malin} \& {Hadley}(1997)}]{Malin97}
{Malin} D., {Hadley} B., 1997, Publications of the Astronomical Society of
  Australia, 14, 52

\bibitem[{{Maraston}(2005)}]{Maraston05}
{Maraston} C., 2005, \mnras, 362, 799

\bibitem[{{Maraston} {et~al.}(2004){Maraston}, {Bastian}, {Saglia},
  {Kissler-Patig}, {Schweizer}, \& {Goudfrooij}}]{Maraston04}
{Maraston} C., {Bastian} N., {Saglia} R.~P., {Kissler-Patig} M., {Schweizer}
  F., {Goudfrooij} P., 2004, \aap, 416, 467

\bibitem[{{McLaughlin} \& {van der Marel}(2005)}]{McLaughlin05}
{McLaughlin} D.~E., {van der Marel} R.~P., 2005, \apjs, 161, 304

\bibitem[{{Meylan} {et~al.}(2001){Meylan}, {Sarajedini}, {Jablonka},
  {Djorgovski}, {Bridges}, \& {Rich}}]{Meylan01}
{Meylan} G., {Sarajedini} A., {Jablonka} P., {Djorgovski} S.~G., {Bridges} T.,
  {Rich} R.~M., 2001, \aj, 122, 830

\bibitem[{{Mieske} {et~al.}(2008{\natexlab{a}}){Mieske}, {Dabringhausen},
  {Kroupa}, {Hilker}, \& {Baumgardt}}]{Mieske08_AN}
{Mieske} S., {Dabringhausen} J., {Kroupa} P., {Hilker} M., {Baumgardt} H.,
  2008{\natexlab{a}}, Astronomische Nachrichten, 329, 964

\bibitem[{{Mieske} {et~al.}(2002){Mieske}, {Hilker}, \& {Infante}}]{Mieske02}
{Mieske} S., {Hilker} M., {Infante} L., 2002, \aap, 383, 823

\bibitem[{{Mieske} {et~al.}(2004){Mieske}, {Hilker}, \& {Infante}}]{Mieske04}
---, 2004, \aap, 418, 445

\bibitem[{{Mieske} {et~al.}(2006{\natexlab{a}}){Mieske}, {Hilker}, {Infante},
  \& {Jord{\'a}n}}]{Mieske06}
{Mieske} S., {Hilker} M., {Infante} L., {Jord{\'a}n} A., 2006{\natexlab{a}},
  \aj, 131, 2442

\bibitem[{{Mieske} {et~al.}(2008{\natexlab{b}}){Mieske}, {Hilker},
  {Jord{\'a}n}, {Infante}, {Kissler-Patig}, {Rejkuba}, {Richtler},
  {C{\^o}t{\'e}}, {Baumgardt}, {West}, {Ferrarese}, \& {Peng}}]{Mieske08}
{Mieske} S., {Hilker} M., {Jord{\'a}n} A., {Infante} L., {Kissler-Patig} M.,
  {Rejkuba} M., {Richtler} T., {C{\^o}t{\'e}} P., {Baumgardt} H., {West} M.~J.,
  {Ferrarese} L., {Peng} E.~W., 2008{\natexlab{b}}, \aap, 487, 921

\bibitem[{{Mieske} {et~al.}(2009){Mieske}, {Hilker}, {Misgeld}, {Jord{\'a}n},
  {Infante}, \& {Kissler-Patig}}]{Mieske09}
{Mieske} S., {Hilker} M., {Misgeld} I., {Jord{\'a}n} A., {Infante} L.,
  {Kissler-Patig} M., 2009, \aap, 498, 705

\bibitem[{{Mieske} {et~al.}(2006{\natexlab{b}}){Mieske}, {Jord{\'a}n},
  {C{\^o}t{\'e}}, {Kissler-Patig}, {Peng}, {Ferrarese}, {Blakeslee}, {Mei},
  {Merritt}, {Tonry}, \& {West}}]{MieskeACS06}
{Mieske} S., {Jord{\'a}n} A., {C{\^o}t{\'e}} P., {Kissler-Patig} M., {Peng}
  E.~W., {Ferrarese} L., {Blakeslee} J.~P., {Mei} S., {Merritt} D., {Tonry}
  J.~L., {West} M.~J., 2006{\natexlab{b}}, \apj, 653, 193

\bibitem[{{Mieske} {et~al.}(2010){Mieske}, {Jordan}, {Cote}, {Peng},
  {Ferrarese}, {Blakeslee}, {Mei}, {Baumgardt}, {Tonry}, {Infante}, \&
  {West}}]{Mieske10}
{Mieske} S., {Jordan} A., {Cote} P., {Peng} E., {Ferrarese} L., {Blakeslee} J.,
  {Mei} S., {Baumgardt} H., {Tonry} J., {Infante} L., {West} M., 2010, ArXiv
  e-prints

\bibitem[{{Misgeld} {et~al.}(2008){Misgeld}, {Mieske}, \& {Hilker}}]{Misgeld08}
{Misgeld} I., {Mieske} S., {Hilker} M., 2008, \aap, 486, 697

\bibitem[{{Murray}(2009)}]{Murray09}
{Murray} N., 2009, \apj, 691, 946

\bibitem[{{Norris} {et~al.}(2008){Norris}, {Sharples}, {Bridges}, {Gebhardt},
  {Forbes}, {Proctor}, {Raul Faifer}, {Carlos Forte}, {Beasley}, {Zepf}, \&
  {Hanes}}]{Norris08}
{Norris} M.~A., {Sharples} R.~M., {Bridges} T., {Gebhardt} K., {Forbes} D.~A.,
  {Proctor} R., {Raul Faifer} F., {Carlos Forte} J., {Beasley} M.~A., {Zepf}
  S.~E., {Hanes} D.~A., 2008, \mnras, 385, 40

\bibitem[{{Norris} {et~al.}(2006){Norris}, {Sharples}, \& {Kuntschner}}]{NSK06}
{Norris} M.~A., {Sharples} R.~M., {Kuntschner} H., 2006, \mnras, 367, 815

\bibitem[{{Paudel} {et~al.}(2010{\natexlab{a}}){Paudel}, {Lisker}, \&
  {Janz}}]{Paudel10b}
{Paudel} S., {Lisker} T., {Janz} J., 2010{\natexlab{a}}, ArXiv e-prints

\bibitem[{{Paudel} {et~al.}(2010{\natexlab{b}}){Paudel}, {Lisker},
  {Kuntschner}, {Grebel}, \& {Glatt}}]{Paudel10a}
{Paudel} S., {Lisker} T., {Kuntschner} H., {Grebel} E.~K., {Glatt} K.,
  2010{\natexlab{b}}, \mnras, 405, 800

\bibitem[{{Peacock} {et~al.}(2009){Peacock}, {Maccarone}, {Waters}, {Kundu},
  {Zepf}, {Knigge}, \& {Zurek}}]{Peacock09}
{Peacock} M.~B., {Maccarone} T.~J., {Waters} C.~Z., {Kundu} A., {Zepf} S.~E.,
  {Knigge} C., {Zurek} D.~R., 2009, \mnras, 392, L55

\bibitem[{{Peng} {et~al.}(2002){Peng}, {Ho}, {Impey}, \& {Rix}}]{Peng02_Galfit}
{Peng} C.~Y., {Ho} L.~C., {Impey} C.~D., {Rix} H.-W., 2002, \aj, 124, 266

\bibitem[{{Peng} {et~al.}(2009){Peng}, {Jord{\'a}n}, {Blakeslee}, {Mieske},
  {C{\^o}t{\'e}}, {Ferrarese}, {Harris}, {Madrid}, \& {Meurer}}]{Peng09}
{Peng} E.~W., {Jord{\'a}n} A., {Blakeslee} J.~P., {Mieske} S., {C{\^o}t{\'e}}
  P., {Ferrarese} L., {Harris} W.~E., {Madrid} J.~P., {Meurer} G.~R., 2009,
  \apj, 703, 42

\bibitem[{{Peng} {et~al.}(2006){Peng}, {Jord{\'a}n}, {C{\^o}t{\'e}},
  {Blakeslee}, {Ferrarese}, {Mei}, {West}, {Merritt}, {Milosavljevi{\'c}}, \&
  {Tonry}}]{Peng06}
{Peng} E.~W., {Jord{\'a}n} A., {C{\^o}t{\'e}} P., {Blakeslee} J.~P.,
  {Ferrarese} L., {Mei} S., {West} M.~J., {Merritt} D., {Milosavljevi{\'c}} M.,
  {Tonry} J.~L., 2006, \apj, 639, 95

\bibitem[{{Phillipps} {et~al.}(2001){Phillipps}, {Drinkwater}, {Gregg}, \&
  {Jones}}]{Phillips01}
{Phillipps} S., {Drinkwater} M.~J., {Gregg} M.~D., {Jones} J.~B., 2001, \apj,
  560, 201

\bibitem[{{Price} {et~al.}(2009){Price}, {Phillipps}, {Huxor}, {Trentham},
  {Ferguson}, {Marzke}, {Hornschemeier}, {Goudfrooij}, {Hammer}, {Tully},
  {Chiboucas}, {Smith}, {Carter}, {Merritt}, {Balcells}, {Erwin}, \&
  {Puzia}}]{Price09}
{Price} J., {Phillipps} S., {Huxor} A., {Trentham} N., {Ferguson} H.~C.,
  {Marzke} R.~O., {Hornschemeier} A., {Goudfrooij} P., {Hammer} D., {Tully}
  R.~B., {Chiboucas} K., {Smith} R.~J., {Carter} D., {Merritt} D., {Balcells}
  M., {Erwin} P., {Puzia} T.~H., 2009, \mnras, 397, 1816

\bibitem[{{Proctor} {et~al.}(2004){Proctor}, {Forbes}, \&
  {Beasley}}]{Proctor04}
{Proctor} R.~N., {Forbes} D.~A., {Beasley} M.~A., 2004, \mnras, 355, 1327

\bibitem[{{Puzia} {et~al.}(2005){Puzia}, {Kissler-Patig}, {Thomas}, {Maraston},
  {Saglia}, {Bender}, {Goudfrooij}, \& {Hempel}}]{Puzia05}
{Puzia} T.~H., {Kissler-Patig} M., {Thomas} D., {Maraston} C., {Saglia} R.~P.,
  {Bender} R., {Goudfrooij} P., {Hempel} M., 2005, \aap, 439, 997

\bibitem[{{Rejkuba} {et~al.}(2007){Rejkuba}, {Dubath}, {Minniti}, \&
  {Meylan}}]{Rejkuba07}
{Rejkuba} M., {Dubath} P., {Minniti} D., {Meylan} G., 2007, \aap, 469, 147

\bibitem[{{Rhode} \& {Zepf}(2004)}]{Rhode04}
{Rhode} K.~L., {Zepf} S.~E., 2004, \aj, 127, 302

\bibitem[{{Robin} {et~al.}(2003){Robin}, {Reyl{\'e}}, {Derri{\`e}re}, \&
  {Picaud}}]{Robin03}
{Robin} A.~C., {Reyl{\'e}} C., {Derri{\`e}re} S., {Picaud} S., 2003, \aap, 409,
  523

\bibitem[{{Sage} \& {Galletta}(1994)}]{Sage94}
{Sage} L.~J., {Galletta} G., 1994, \aj, 108, 1633

\bibitem[{{S{\'a}nchez-Bl{\'a}zquez} {et~al.}(2006){S{\'a}nchez-Bl{\'a}zquez},
  {Gorgas}, {Cardiel}, \& {Gonz{\'a}lez}}]{SanchezBlazquez06b}
{S{\'a}nchez-Bl{\'a}zquez} P., {Gorgas} J., {Cardiel} N., {Gonz{\'a}lez} J.~J.,
  2006, \aap, 457, 809

\bibitem[{{Sarzi} {et~al.}(2006){Sarzi}, {Falc{\'o}n-Barroso}, {Davies},
  {Bacon}, {Bureau}, {Cappellari}, {de Zeeuw}, {Emsellem}, {Fathi},
  {Krajnovi{\'c}}, {Kuntschner}, {McDermid}, \& {Peletier}}]{Sauron5}
{Sarzi} M., {Falc{\'o}n-Barroso} J., {Davies} R.~L., {Bacon} R., {Bureau} M.,
  {Cappellari} M., {de Zeeuw} P.~T., {Emsellem} E., {Fathi} K., {Krajnovi{\'c}}
  D., {Kuntschner} H., {McDermid} R.~M., {Peletier} R.~F., 2006, \mnras, 366,
  1151

\bibitem[{{Schlegel} {et~al.}(1998){Schlegel}, {Finkbeiner}, \&
  {Davis}}]{Schlegel1998}
{Schlegel} D.~J., {Finkbeiner} D.~P., {Davis} M., 1998, \apj, 500, 525

\bibitem[{{Sersic}(1968)}]{Sersic68}
{Sersic} J.~L., 1968, {Atlas de galaxias australes}. Cordoba

\bibitem[{{Sikkema} {et~al.}(2006){Sikkema}, {Peletier}, {Carter}, {Valentijn},
  \& {Balcells}}]{Sikkema06}
{Sikkema} G., {Peletier} R.~F., {Carter} D., {Valentijn} E.~A., {Balcells} M.,
  2006, \aap, 458, 53

\bibitem[{{Sirianni} {et~al.}(2005){Sirianni}, {Jee}, {Ben{\'{\i}}tez},
  {Blakeslee}, {Martel}, {Meurer}, {Clampin}, {De Marchi}, {Ford}, {Gilliland},
  {Hartig}, {Illingworth}, {Mack}, \& {McCann}}]{Sirianni05}
{Sirianni} M., {Jee} M.~J., {Ben{\'{\i}}tez} N., {Blakeslee} J.~P., {Martel}
  A.~R., {Meurer} G., {Clampin} M., {De Marchi} G., {Ford} H.~C., {Gilliland}
  R., {Hartig} G.~F., {Illingworth} G.~D., {Mack} J., {McCann} W.~J., 2005,
  \pasp, 117, 1049

\bibitem[{{Skrutskie} {et~al.}(2006){Skrutskie}, {Cutri}, {Stiening}, \& {et
  al.,}}]{Skrutskie06}
{Skrutskie} M.~F., {Cutri} R.~M., {Stiening} R., {et al.,}, 2006, \aj, 131,
  1163

\bibitem[{{Spitler} {et~al.}(2006){Spitler}, {Larsen}, {Strader}, {Brodie},
  {Forbes}, \& {Beasley}}]{Spitler06}
{Spitler} L.~R., {Larsen} S.~S., {Strader} J., {Brodie} J.~P., {Forbes} D.~A.,
  {Beasley} M.~A., 2006, \aj, 132, 1593

\bibitem[{{Strader} {et~al.}(2006){Strader}, {Brodie}, {Spitler}, \&
  {Beasley}}]{Strader06}
{Strader} J., {Brodie} J.~P., {Spitler} L., {Beasley} M.~A., 2006, \aj, 132,
  2333

\bibitem[{{Tamura} {et~al.}(2006){Tamura}, {Sharples}, {Arimoto}, {Onodera},
  {Ohta}, \& {Yamada}}]{Tamura06}
{Tamura} N., {Sharples} R.~M., {Arimoto} N., {Onodera} M., {Ohta} K., {Yamada}
  Y., 2006, \mnras, 373, 601

\bibitem[{{Taylor} {et~al.}(2010){Taylor}, {Puzia}, {Harris}, {Harris},
  {Kissler-Patig}, \& {Hilker}}]{Taylor10}
{Taylor} M.~A., {Puzia} T.~H., {Harris} G.~L., {Harris} W.~E., {Kissler-Patig}
  M., {Hilker} M., 2010, \apj, 712, 1191

\bibitem[{{Thomas} {et~al.}(2003){Thomas}, {Maraston}, \& {Bender}}]{Thomas03}
{Thomas} D., {Maraston} C., {Bender} R., 2003, \mnras, 339, 897

\bibitem[{{Thomas} {et~al.}(2004){Thomas}, {Maraston}, \& {Korn}}]{Thomas04}
{Thomas} D., {Maraston} C., {Korn} A., 2004, \mnras, 351, L19

\bibitem[{{Tonry} {et~al.}(2001){Tonry}, {Dressler}, {Blakeslee}, {Ajhar},
  {Fletcher}, {Luppino}, {Metzger}, \& {Moore}}]{TonrySBF4}
{Tonry} J.~L., {Dressler} A., {Blakeslee} J.~P., {Ajhar} E.~A., {Fletcher}
  A.~B., {Luppino} G.~A., {Metzger} M.~R., {Moore} C.~B., 2001, \apj, 546, 681

\bibitem[{{Trager} {et~al.}(1998){Trager}, {Worthey}, {Faber}, {Burstein}, \&
  {Gonzalez}}]{Trager98}
{Trager} S.~C., {Worthey} G., {Faber} S.~M., {Burstein} D., {Gonzalez} J.~J.,
  1998, \apjs, 116, 1

\bibitem[{{Vazdekis} {et~al.}(2010){Vazdekis}, {S{\'a}nchez-Bl{\'a}zquez},
  {Falc{\'o}n-Barroso}, {Cenarro}, {Beasley}, {Cardiel}, {Gorgas}, \&
  {Peletier}}]{Vazdekis10}
{Vazdekis} A., {S{\'a}nchez-Bl{\'a}zquez} P., {Falc{\'o}n-Barroso} J.,
  {Cenarro} A.~J., {Beasley} M.~A., {Cardiel} N., {Gorgas} J., {Peletier}
  R.~F., 2010, \mnras, 404, 1639

\bibitem[{{Wehner} \& {Harris}(2007)}]{Wehner07}
{Wehner} E.~M.~H., {Harris} W.~E., 2007, \apjl, 668, L35

\bibitem[{{Wei} {et~al.}(2010){Wei}, {Kannappan}, {Vogel}, \& {Baker}}]{Wei10}
{Wei} L.~H., {Kannappan} S.~J., {Vogel} S.~N., {Baker} A.~J., 2010, \apj, 708,
  841

\end{thebibliography}

\begin{appendix}
\section{Investigation of the NGC3923 Blue Tilt}
\label{sec:Appendix_Blue_Tilt}

A closer look at Figure \ref{fig:ngc3923_cm} indicates the possible presence of a ``blue tilt'' 
\citep{Harris06,MieskeACS06,Strader06} in the blue GC population. The blue tilt appears to 
be a common feature of the GC systems of massive galaxies and is usually interpreted as 
being the result of a mass-metallicity relation due to more massive GCs being able to self-enrich 
more than their lower mass counterparts.

To determine the significance of the NGC3923 blue tilt we follow a procedure similar to that
used by most other studies of this phenomenon \citep[e.g. ][for M87 GCs]{Peng09}. We limit 
our sample to GCs with M$_{\rm I}$ $<$ --8.34, ensuring that we work only with objects with 
acceptable photometric errors. The effect of this limit is likely to be negligible, as the blue tilt 
is found to be driven by GCs more luminous than M$_{\rm I}$ $\sim$ --10 \citep{Harris09,Peng09}. 
We next bin our GCs by magnitude into four bins of approximately equal numbers of GCs 
(N $>$ 100 in each). We fit the peak positions of the blue and red GC distributions of each bin 
using the KMM method \citep{KMM}, with errors in the measured colours of the red and blue 
peaks determined from 50 Monte-Carlo resimulations of the data within the photometric noise. 
We then fit a simple linear relation to the positions of the red and blue peaks. This is likely to 
be a significant simplification as the blue tilt is expected to display a curve at higher luminosities 
\citep[see][]{Harris09}. However with the limited statistics available for this sample a first order 
model is all that can be reliably fit.

The best-fit slopes $\gamma_{\rm I}=\frac{d(V-I)}{dI}$ to the peak colours of the blue and red 
populations are $\gamma$$_{\rm I,blue}$= --0.024 $\pm$ 0.007 and $\gamma$$_{\rm I,red}$ =
+0.012 $\pm$ 0.013 respectively. We therefore detect a blue tilt at greater than 3$\sigma$ 
significance. However, in common with other studies we find no compelling evidence for a 
corresponding trend in the red GCs of NGC3923. Our measured slopes are indistinguishable 
from those measured by \citet{Peng09} for the GCs of M87 ($\gamma$$_{\rm I,blue}$= --0.024 
$\pm$ 0.006 and $\gamma$$_{\rm I,red}$ = +0.003 $\pm$ 0.007).

\section{Investigation of the NGC3923 GCLF}
\label{sec:Appendix_GCLF}

Figure \ref{fig:NGC3923_GCLF} shows the results of our examination of the GCLF of NGC3923. 
In this analysis we have made use of the photometry of all GCs examined in Sections \ref{Sec:HST_IP} 
and \ref{Sec:ngc3923gcs} with V $<$ 24.75, ensuring we are operating within the region where 
incompleteness is negligible (the I band 90$\%$ and 50$\%$ completeness limits are 24.3 and 
24.6 respectively). We remove contamination following the methods of \cite{Sikkema06}. To 
statistically remove foreground stars we make use of the Besancon model for the distribution of 
Milky Way stars \citep{Robin03}. Unresolved background galaxies were removed statistically by 
applying our GC selection criteria to observations of the Hubble Deep Field South.

We bin the resulting V-band GC photometry and fit a single Gaussian to the observed histogram.  
As shown in Figure \ref{fig:NGC3923_GCLF} the resulting turnover magnitude is entirely consistent 
with the observed universal GCLF turnover magnitude of $\sim$~--7.4~$\pm$~0.1 
\citep[e.g.][]{Kundu01a,Kundu01b,Jordan07} for the assumed distance of NGC3923. The GCLF 
analysis therefore provides independent verification of the assumed distance modulus of NGC3923. 
The measured 1.27~mag GCLF dispersion of the NGC3923 GCLF is also consistent with values 
found for other massive ellipticals \citep[see e.g.][]{Hilker09}.

\end{appendix}

\label{lastpage}
\end{document}